\documentclass[11pt]{article}
\usepackage[centertags]{amsmath}
\usepackage{amsfonts}
\usepackage{eurosym}
\usepackage{amssymb}
\usepackage{amsthm}
\usepackage{amsmath}
\usepackage{version}
\usepackage{multirow}
\usepackage{tabularx}
\usepackage{bbm}
\usepackage{algorithm}
\usepackage{algorithmic}
\usepackage{placeins}
\usepackage{subfigure}
\usepackage{slashbox}
\usepackage[english]{babel}
\usepackage{latexsym,graphicx}
\usepackage{color}

\usepackage{geometry}
\theoremstyle{plain}
\newtheorem{thm}{Theorem}[section]
\newtheorem{cor}[thm]{Corollary}

\newtheorem{prop}[thm]{Proposition}

\theoremstyle{definition}

\theoremstyle{remark}
\newtheorem{rem}{\bf Remark}[section]
\theoremstyle{remark}

\theoremstyle{remark}
\newtheorem{example}{\bf Example}[section]
\usepackage{fancyhdr}

\DeclareMathOperator*{\argmin}{arg\,min}

\newcommand{\NIG}{\textup{NIG}}
\newcommand{\blackboard}[1]{\mathbf{#1}} 
\newcommand{\R}{\blackboard{R}}
\renewcommand{\P}{\blackboard{P}}
\newcommand{\E}{\blackboard{E}}

\DeclareMathOperator{\var}{var}
\DeclareMathOperator{\corr}{cor}

\renewcommand{\d}[1]{\textup{d} #1}

\newcommand{\delimleft}[2]{\ifcase #1\or
    \bigl#2\or %
    \Bigl#2\or %
    \biggl#2\or %
    \Biggl#2\or %
    \left#2\fi}
\newcommand{\delimright}[2]{\ifcase #1\or
    \bigr#2\or %
    \Bigr#2\or %
    \biggr#2\or %
    \Biggr#2\or %
    \right#2\fi}
\newcommand{\pa}[2][5]{
    \delimleft{#1}{(} #2 \delimright{#1}{)}}
\newcommand{\br}[2][5]{
    \delimleft{#1}{[} #2 \delimright{#1}{]}}
\newcommand{\ac}[2][5]{
    \delimleft{#1}{\{} #2 \delimright{#1}{\}}}

\newcommand{\prob}[2][5]{
    \P \br[#1]{#2}}

\newcommand{\esp}[2][5]{
    \E \br[#1]{#2}}

\newcommand{\proc}[2][0]{
    {\bigl( #2_t \bigr)}_{t \ge #1}}

\newcommand{\ie}{\textit{i.e. }}
\renewcommand{\le}{\leqslant}
\renewcommand{\ge}{\geqslant}

\begin{document}
\title{Joint Modelling of Gas and Electricity spot prices}
\author{{\normalsize N. Frikha${}^{1}$ , V. Lemaire${}^{2}$}} 
\date{\today}
\maketitle
\begin{abstract}
    The recent liberalization of the electricity and gas markets has resulted in the growth of energy exchanges and modelling problems. In this paper, we modelize jointly gas and electricity spot prices using a mean-reverting model which fits the correlations structures for the two commodities. The dynamics are based on Ornstein processes with parameterized diffusion coefficients. Moreover, using the empirical distributions of the spot prices, we derive a class of such parameterized diffusions which captures the most salient statistical properties: stationarity, spikes and heavy-tailed distributions. 

    The associated calibration procedure is based on standard and efficient statistical tools. We calibrate the model on French market for electricity and on UK market for gas, and then we simulate some trajectories which reproduce well the observed prices behavior. Finally, we illustrate the importance of the correlation structure and of the presence of spikes by measuring the risk on a power plant portfolio.
\end{abstract}
   
\footnotetext[{1}]{Laboratoire de Probabilit\'{e}s et Mod\`{e}les al\'{e}atoires, UMR 7599, Universit\'{e} Pierre et Marie Curie, France, e-mail: frikha.noufel@gmail.com}
\footnotetext[{2}]{Laboratoire de Probabilit\'{e}s et Mod\`{e}les al\'{e}atoires, UMR 7599, Universit\'{e} Pierre et Marie Curie, France, e-mail: vincent.lemaire@upmc.fr}

\section{Introduction}
The recent deregulation of energy markets has led to the development in several countries of market places for energy exchanges. Consequently, understanding and modelling the behavior of energy market is necessary for developing a risk management framework as well as pricing of options. Many derivatives on both electricity and gas spot (and futures) prices are traded. Understanding the correlation structure between both energies is a significant challenge. For instance, spark spread options are commonly traded in energy markets as a way to hedge price differences between electricity and gas prices or are used in order to price projects in energy (see \cite{geman} for an introduction). Thus, modelling jointly the evolution of gas and electricity prices is a relevant issue. 
 
Numerous diffusion-type and econometric models have been proposed for electricity and gas spot prices. In energy markets, spot price dynamics are commonly based on Ornstein processes, which are the classical way to model mean-reversion. Geometric models represent the logarithmic prices by a sum of Ornstein processes with different speeds of mean reversion whereas arithmetic models represent the price itself (see for instance \cite{schwartz} for a geometric model). Also, equilibrium models (\cite{barlow} and \cite{kanamura}) have been investigated in order to reproduce price formation as a balance between supply and demand. The main drawback of such model is that they do not reproduce the autocorrelation structure of a commodity and the cross-correlation structure between commodities. In \cite{gemanroncorini}, a markov jump diffusion is investigated for electricity spot prices. Though, it properly represents the spiky behaviour of spot electricity prices, the process reverts to a deterministic mean level whereas it usually reverts to the pre-spike value on data. Moreover applied to electricity and gas spot prices, it does not capture the autocorrelation and cross-correlation struture observed on data. 

Another class of spot price dynamics is represented by multifactor models. Several authors (see \cite{Hambly2009}, \cite{benth}, \cite{deng}, \cite{tankov}, \cite{villaplana} among others) have investigated this kind of diffusion. The logarithmic prices or the price itself is represented by a sum of Ornstein processes in order to incorporate a mixture of jump variations and ``normal'' variations. For instance, in \cite{tankov} the deseasonalized spot price or log-spot price $X(t)$ is given by:
$$X(t)=Y_{1}(t)+Y_{2}(t)$$ 
where 
$$
\mbox{d}Y_{i}(t)=-\lambda_{i}Y_{i}(t) \mbox{d}t +\mbox{d}L_{i}(t), \ \ i=1,2.
$$

\noindent The Ornstein Uhlenbeck (OU) component $Y_{1}$ is responsible for the normal variation and is assumed to be Gaussian, \ie $L_{1}(t)$ is a Brownian motion, whereas $Y_{2}$ is the Levy driven OU component responsible for spikes, \ie $L_{2}(t)$ is a jump L\'{e}vy process. In this kind of framework, the difficulty is to detect and filter the spikes in order to estimate the jump part. Several methods have been proposed to circumvent this problem (see e.g. \cite{tankov} and \cite{benth}). For instance, \cite{Hambly2009} presents a similar model for the spot price process that is the exponential of the sum of an Ornstein-Uhlenbeck and an independent mean reverting pure jump process, derives its associated forward curves and finally proposes to calibrate it to the observed forward curve at time $t=0$. In order to calculate premia of call and put options as well as path-dependent options several approximations to the probability density function of the logarithm of the spot price process at maturity are done. In \cite{benthkufakunuseu}, the following spot price dynamics for two energies $A$ and $B$ are proposed 
\begin{align*}
S^{A}(t) &= \Lambda^{A}(t)+ \sum_{i=1}^{m} X_{i}^{A}(t) + \sum_{j=1}^{n} Y_{j}^{A}(t), \\
S^{B}(t) &= \Lambda^{B}(t)+ \sum_{i=1}^{m} X_{i}^{B}(t) + \sum_{j=1}^{n} Y_{j}^{B}(t),
\end{align*}

\noindent where $\Lambda^{A}(t)$ and $\Lambda^{B}(t)$ are seasonal floors, $X_{i}^{A}$ and $X_{i}^{B}$ are common OU processes, $i.e.$ they are driven by the same jump process $L_{i}$. A different approach based on copula is proposed in \cite{benthkettler} where the joint evolution of electricity and gas prices is modeled by a bivariate non-Gaussian OU pure jump process with a non-symmetric copula.  

In this paper, we propose an alternative class (arithmetic and geometric) of models to reproduce adequately the statistical features of gas and electricity spot prices based on parameterized local volatility processes. The spiky behaviour of both spot prices is captured \emph{without introducing jump diffusion processes}. More precisely, the deseasonalized (log) prices processes are modelized by the sum of an Ornstein-Uhlenbeck process (which is common for both commodities) and an independent stationary diffusion process. The construction of this stationary diffusion is similar to \cite{sorensen} where diffusion models with linear drift and prespecified marginal distribution are investigated with an application in a different context. The selected parameterized diffusion coefficient allows to capture ``cluster of volatility'' (e.g. period of high volatility which implies prices spikes). \\
Moreover, this approach provides a significant advantage over the class of jump diffusion models since the calibration process involves only classical statistical tools like least squares method and maximum likellihood estimations so that it is robust and fast. It allows to reproduce (for the first time to our knowledge) both the auto-correlation and the cross-correlation strutures between two energies. The model was successfully tested on several markets and seems to fit well the statistical features and the marginal distributions of gas and electricity spot prices.

Our results are presented as follows. Section 2 is devoted to the description of the stylised features of gas and electricity spot prices. Then, in Section 3, we briefly recall some important theoretical results on which are based our model. To be more precise, we recall how to construct a mean reverting diffusion process $X$ solution of a stochastic differential equation (SDE) with a prespecified continuous invariant density $f$. Such diffusions involves parameterized local volatility processes. In Section 4, we present the model of our choice and focus on the calibration procedure. In the last section, we perform the calibration on the data sets coming from the NBP for the gas spot price and the Powernext market for the electricity spot price. Then, we proceed to the simulation and, finally, analyze the impact of the modelization by measuring the risk of an energy related portfolio using several models. We show that introducing the cross-commodity correlation structure can greatly modify the risk of a portfolio.
 
\section{Stylised features of gas and electricity spot prices} 

\subsection{Seasonality} \label{sect-seaso}
\begin{figure}[!ht]
\begin{center}
\includegraphics[width=15cm,height=8cm]{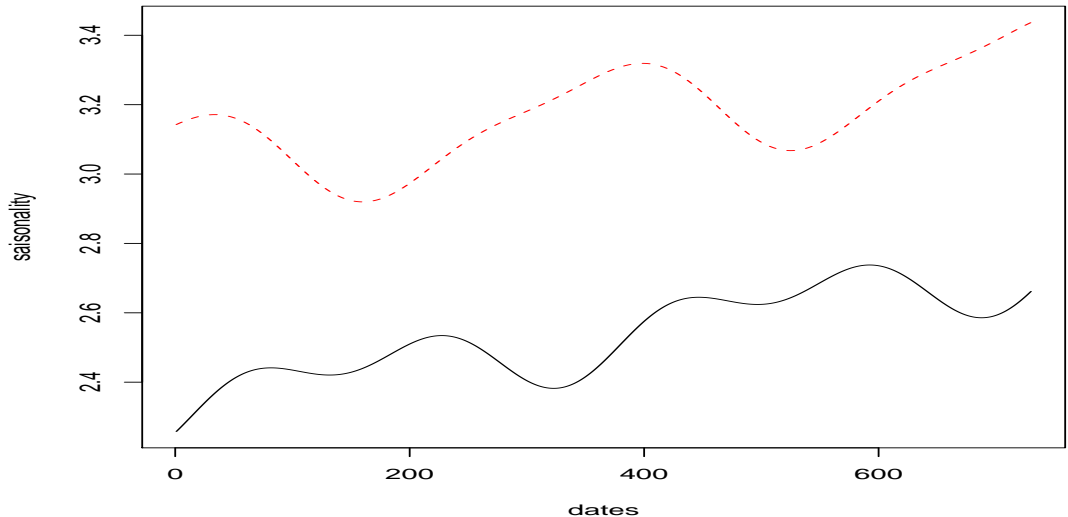}
\caption{The fitted log-seasonality functions $\log(g(t))$ and $\log(e(t))$}
\label{fig-saiso}
\end{center}
\end{figure}

A first characteristic of gas and electricity (and many commodities) prices is the presence of annual (and possibly multi-time scales) seasonality and a trend (see e.g. \cite{geman}, \cite{tankov}). For each commodity, we model the seasonality and the trend component of the logarithmic spot prices with the mean level functions around which spot prices fluctuate
\begin{align*}
\log g(t) & = a^{g} + b^{g} t + \sum_{k=1}^{m} c^{g}_{k} \cos\left(\frac{2\pi t}{l_{k}}\right) + d^{g}_{k} \sin\left( \frac{2\pi t}{l_{k}}\right), \\
\log e(t) & = a^{e} + b^{e} t + \sum_{k=1}^{m} c^{e}_{k} \cos\left(\frac{2\pi t}{l_{k}}\right) + d^{e}_{k} \sin\left( \frac{2\pi t}{l_{k}}\right), 
\end{align*}
where $l_{k}=\left\lfloor 252/k\right\rfloor$, $k=1,\ ...,m$, $\left\lfloor x\right\rfloor$ denotes the integer part of $x$. For instance, if we choose $m=2$, we only consider a seasonal function over the year and the semester. We assume 252 trading days in a year except for electricity spot price on Powernext which has 365 trading days in a year so that, we have to take into account \emph{this particularity} in the seasonality function. The coefficients above are estimated using ordinary least squares. The log-seasonality functions are represented with the estimated values for $m=2$ using gas spot price at the NBP and electricity spot price from the Powernext market in Figure \ref{fig-saiso}. All parameters are not significant at the 5\% level. We only report and take into account the significant values \footnote{$a^{g} = 1.53, b^{g}=0.000688, c_{1}^{g}=0.121, d_{2}^{g}=0.0287, c_{2}^{g}=0.00533$ et $a^{e} = 3.02, b^{e}=0.000405, c_{1}^{e}=0.138, d_{2}^{e}=0.0368.$}. We checked the seasonality over week, month and quater, but the coefficients were not significant.

Now we focus our attention on the deseasonalized data $Y^{g}(t):=\log S^{g}(t)-\log g(t)$ and $Y^{e}(t):=\log S^{e}(t)- \log e(t)$ for the specification of the model. A geometric model consists in modelling the stochastic processes $Y^{g}(t)$ and $ Y^{e}(t)$ whereas an arithmetic model consists in modelling the stochastic processes $e^{Y^{g}(t)}$ and $e^{Y^{e}(t)}$.

\subsection{Spikes and heavy tails}
\begin{figure}[!ht]
\begin{center}
\includegraphics[width=8cm,height=6cm]{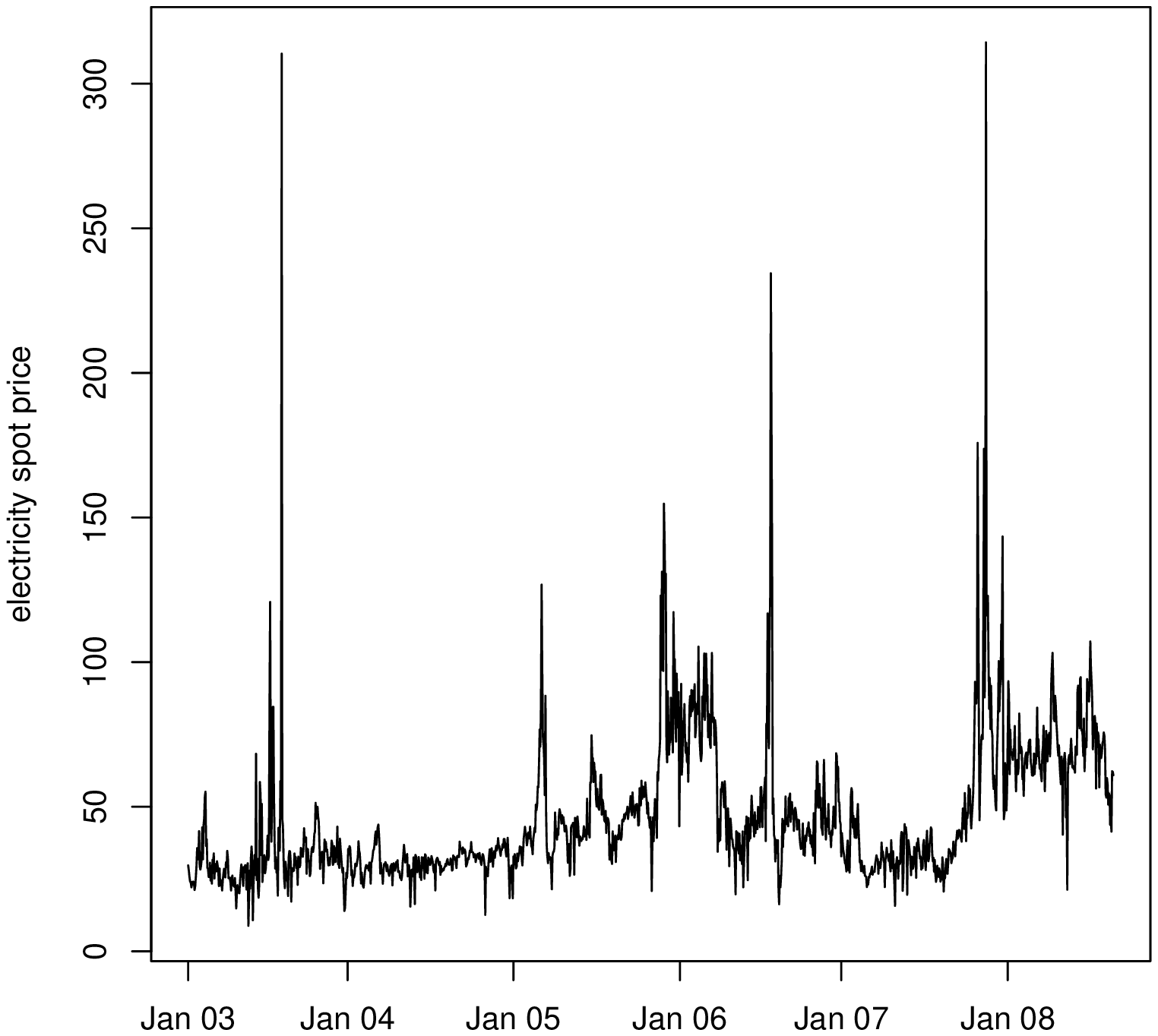}
\includegraphics[width=8cm,height=6cm]{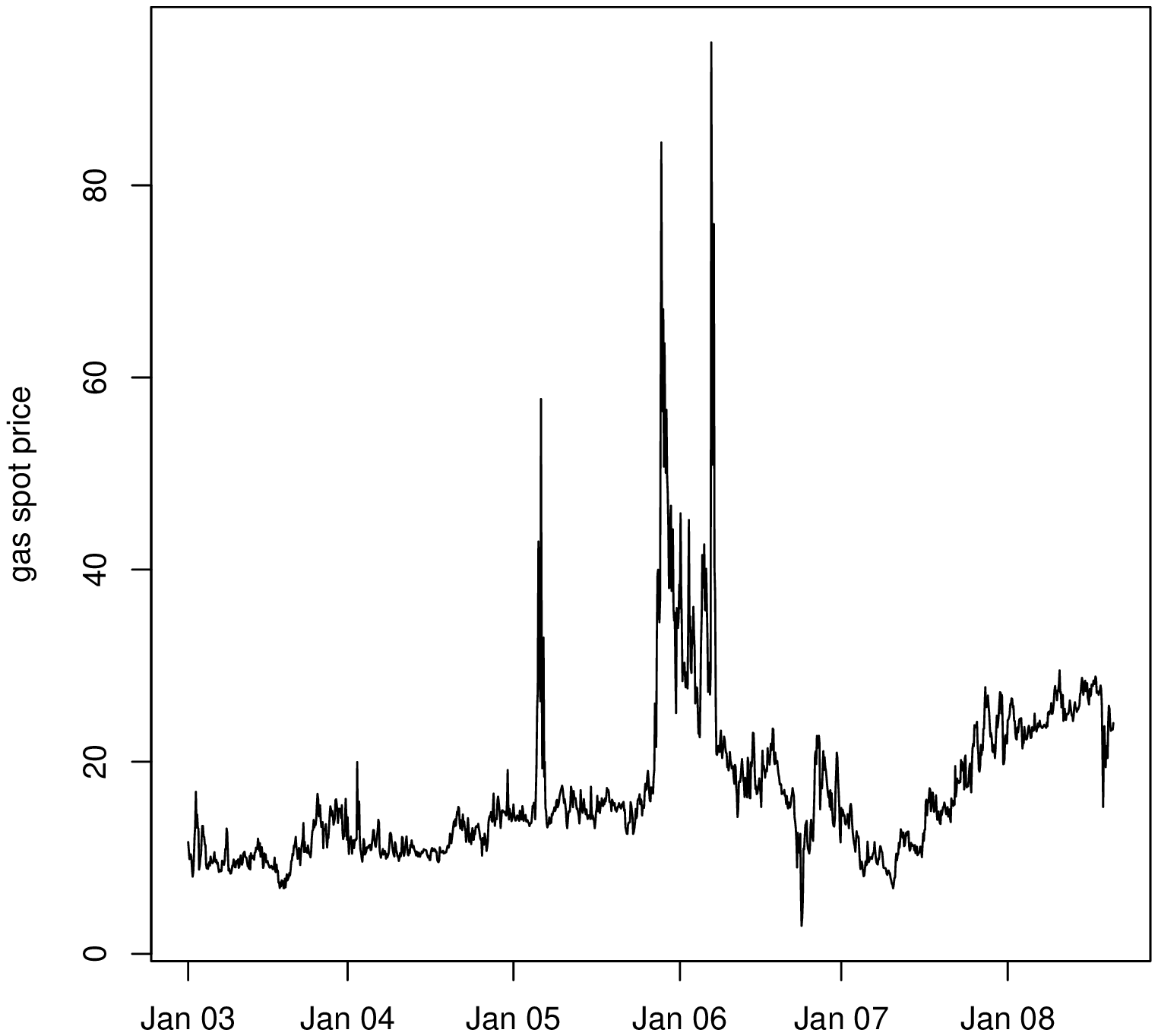}
\caption{Electricity spot prices on the Powernext market (on the left) and gas spot prices at the NBP (on the right) for the period 14 January 2003 till 20 August 2008.}
\label{fig-prix}
\end{center}
\end{figure}

Electricity has very limited storage possibilities. It induces the possibility of spikes in spot prices. Natural gas can be stored but it is often costly, so that it shares the spiky behaviour of spot electricity prices. Gas and electricity markets share this similarity as it can be seen in Figure \ref{fig-prix} presenting the electricity spot prices coming from the Powernext market on the left and gas spot prices at the National Balancing Point (NBP) on the right. From a stochastic modelling point of view, spikes are commonly represented by jump diffusions with mean reversion. However (to the best of our knowledge) there is no evidence that it is rather jumps than spikes caused by clusters of volatility for instance. 

\begin{figure}[!ht]
\begin{center}
\includegraphics[width=8cm,height=6cm]{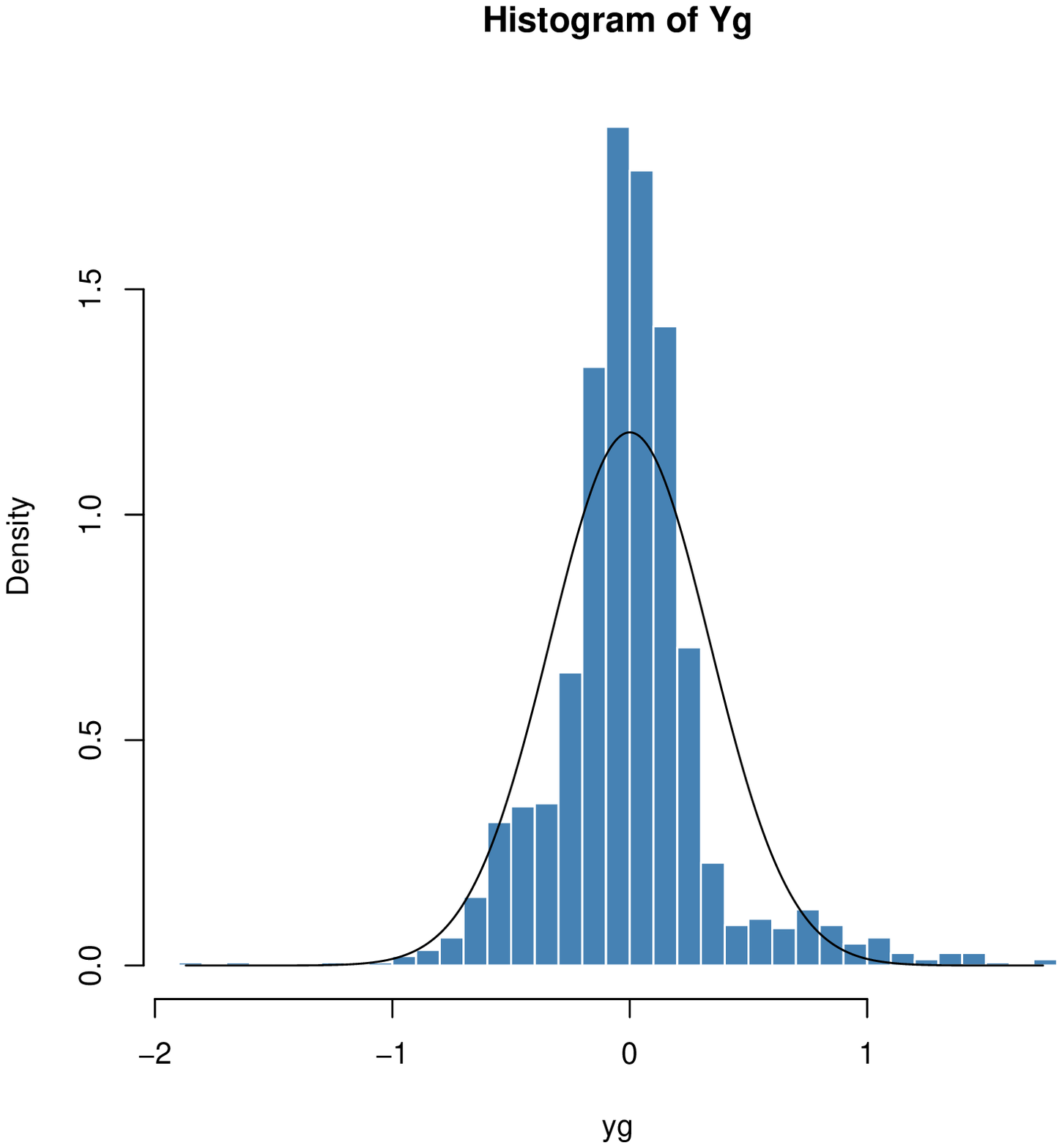}
\includegraphics[width=8cm,height=6cm]{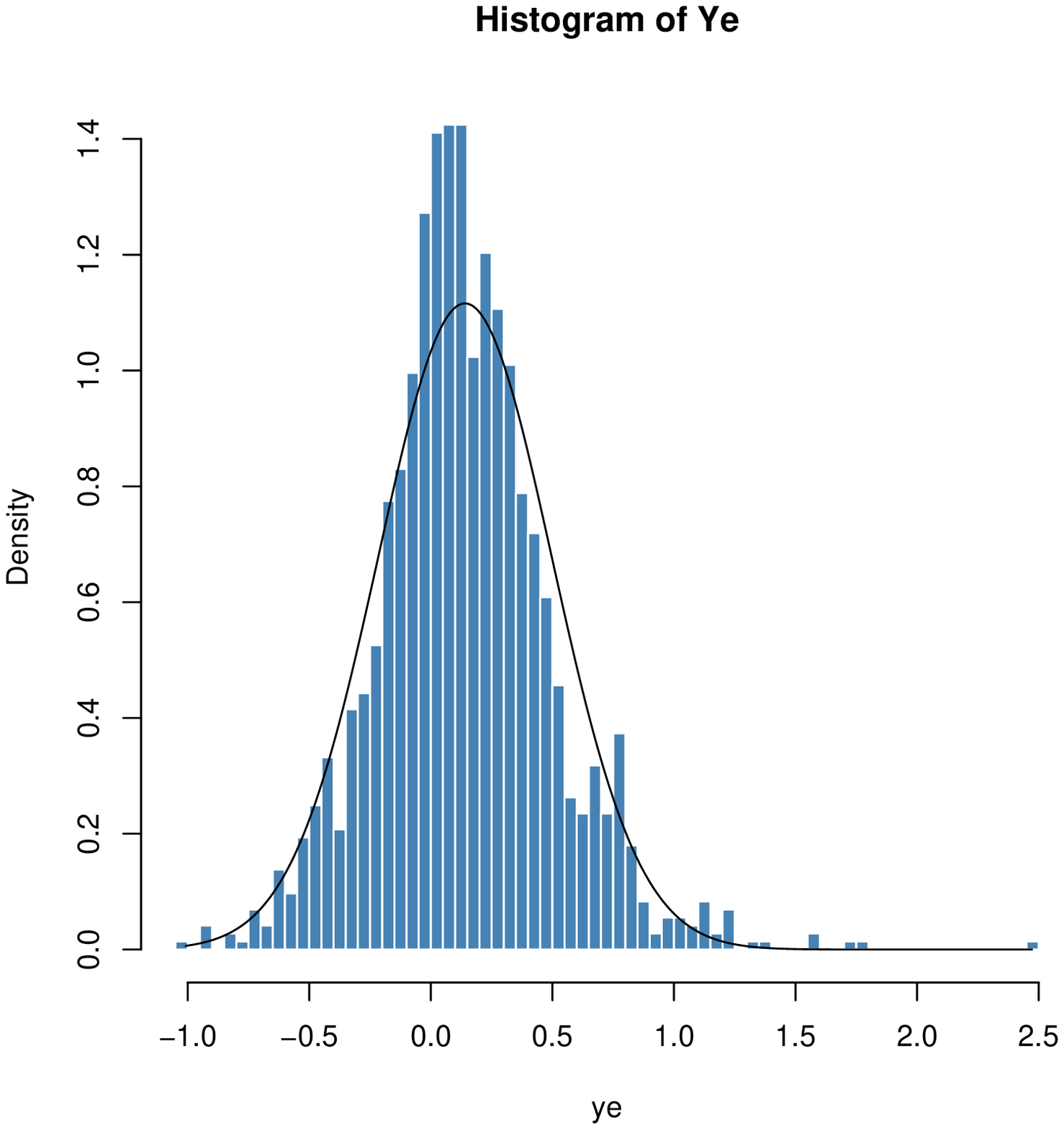}
\caption{Histograms of $Y^{g}$ and $Y^{e}$ with normal density curves.}
\label{fig-histo}
\end{center}
\end{figure}

The histograms of $Y^{g}$ and $Y^{e}$ with the fitted normal density curve is presented in Figure \ref{fig-histo}. We observe that the two residuals time series $Y^{g}$ and $Y^{e}$ are far from being normally distributed. The excess of kurtosis of $Y^{g}$ and $Y^{e}$ are respectively equal to 4.5 and 2.3 meaning that the two distributions are peaked and have heavy tails. The skewness of $Y^{g}$ and $Y^{e}$ are respectively equal to 0.77 and 0.57 meaning that the two distributions are not symmetric.

\subsection{Mean reversion and long term dependency}
\begin{figure}[!ht]
\begin{center}
\includegraphics[width=15cm,height=8cm]{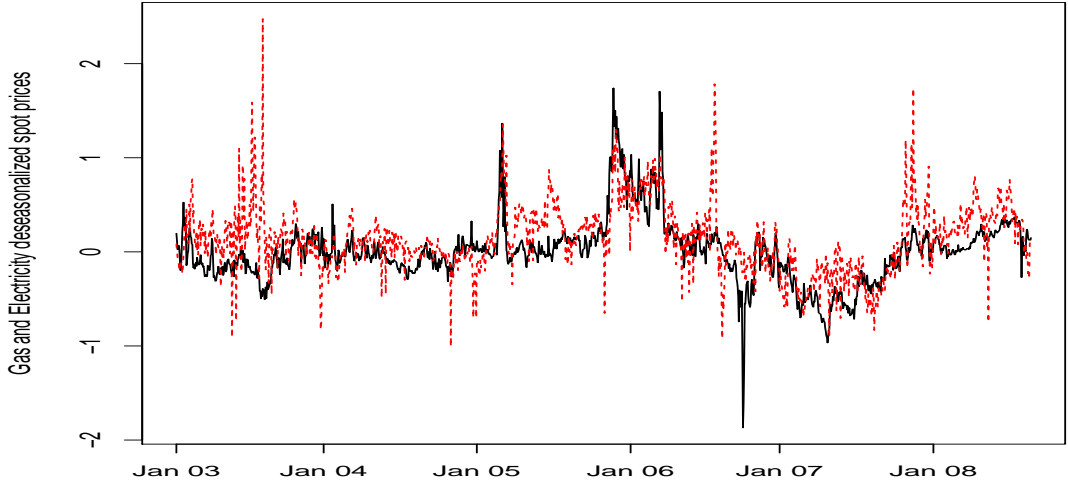}
\caption{The log-deseasonalized gas (normal line) and electricity spot (dashed line) prices}
\label{fig-resid}
\end{center}
\end{figure}

Gas and Electricity spot prices are known to be stationary. This can be tested using an augmented Dickey-Fuller test (ADF) or the Phillips-Perron test. For the UKPX, Powernext electricity spot prices and gas spot prices at the NBP the unit root hypothesis was rejected using both tests. Figure \ref{fig-resid} shows that gas and electricity deseasonalized prices are strongly linked by a long term dependency, \ie it seems that there is a stochastic equilibrium between $Y^{g}(t)$ and $Y^{e}(t)$ from which they cannot deviate for a long time. This long term dependency can be observed on the cross-correlation function.

\subsection{Auto-correlation and cross-correlation}
In energy spot price modelling, the auto-correlation functions (ACFs) are often analyzed. The ACFs of both $Y^{g}(t)$ (respectively $e^{Y^{g}(t)}$), $\rho^{g}$, on one hand $Y^{e}(t)$ (respectively $e^{Y^{e}(t)}$), $\rho^{e}$, on the other hand present both a two-scale (or three-scale at most) decreasing behaviour with one quickly decreasing component and one or two slow decreasing components. The same behaviour is observed on the cross-correlation function (CCF) $\rho^{g,e}$. This kind of decreasing ACFs and CCF are well explained by sum of decreasing exponentials components, namely for $\tau>0$: 
\begin{align*}
\rho^{g}(\tau) & = \textnormal{Corr}\left(Y^{g}(t+\tau), Y^{g}(t)\right)= \phi^{g}_{1}e^{-\lambda^{g}_{1}\tau}+(1-\phi^{g}_{1})e^{-\lambda^{g}_{2}\tau}, \\
\rho^{e}(\tau) & =  \textnormal{Corr}\left(Y^{e}(t+\tau), Y^{e}(t)\right)= \phi^{e}_{1}e^{-\lambda^{e}_{1}\tau}+(1-\phi^{e}_{1})e^{-\lambda^{e}_{2}\tau}, \\
\rho^{g,e}(\tau)& = \textnormal{Corr}\left(Y^{g}(t+\tau), Y^{e}(t)\right)= \phi^{g,e}e^{-\lambda^{g,e}\tau}.
\label{ACFs_CCF} 
\end{align*}

For the sake of simplicity in our stochastic modelization, we focused on one type of cross-correlation $\textnormal{Corr}\left(Y^{g}(t+\tau), Y^{e}(t)\right)$ and we assumed that the cross-correlation is symetric that is $\textnormal{Corr}\left(Y^{g}(t+\tau), Y^{e}(t)\right)=\textnormal{Corr}\left(Y^{e}(t+\tau), Y^{g}(t)\right)$ which is a rather natural approximation. We observed that the slower rates of mean reversion for each commodities are quite similar $\lambda^{g}_{2}=\lambda^{e}_{2}$ and that a rather good approximation is obtained by setting $\lambda^{g,e}=\lambda^{g}_{2}=\lambda^{e}_{2}$. 
\begin{figure}[!ht]
\begin{center}
	\subfigure[ACF of $Y^g$]{\includegraphics[width=5cm,height=5cm]{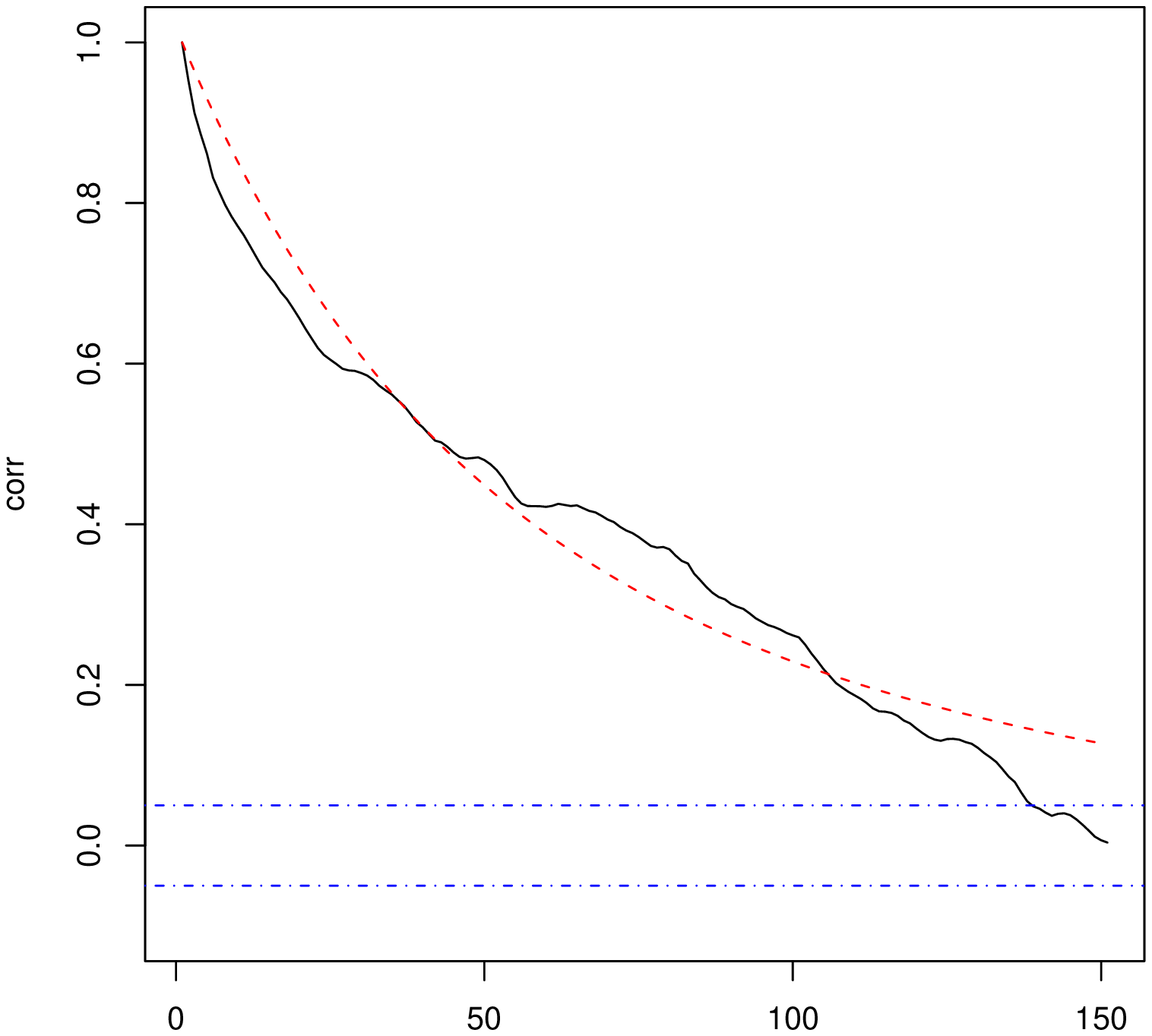}}
	\subfigure[ACF of $Y^e$]{\includegraphics[width=5cm,height=5cm]{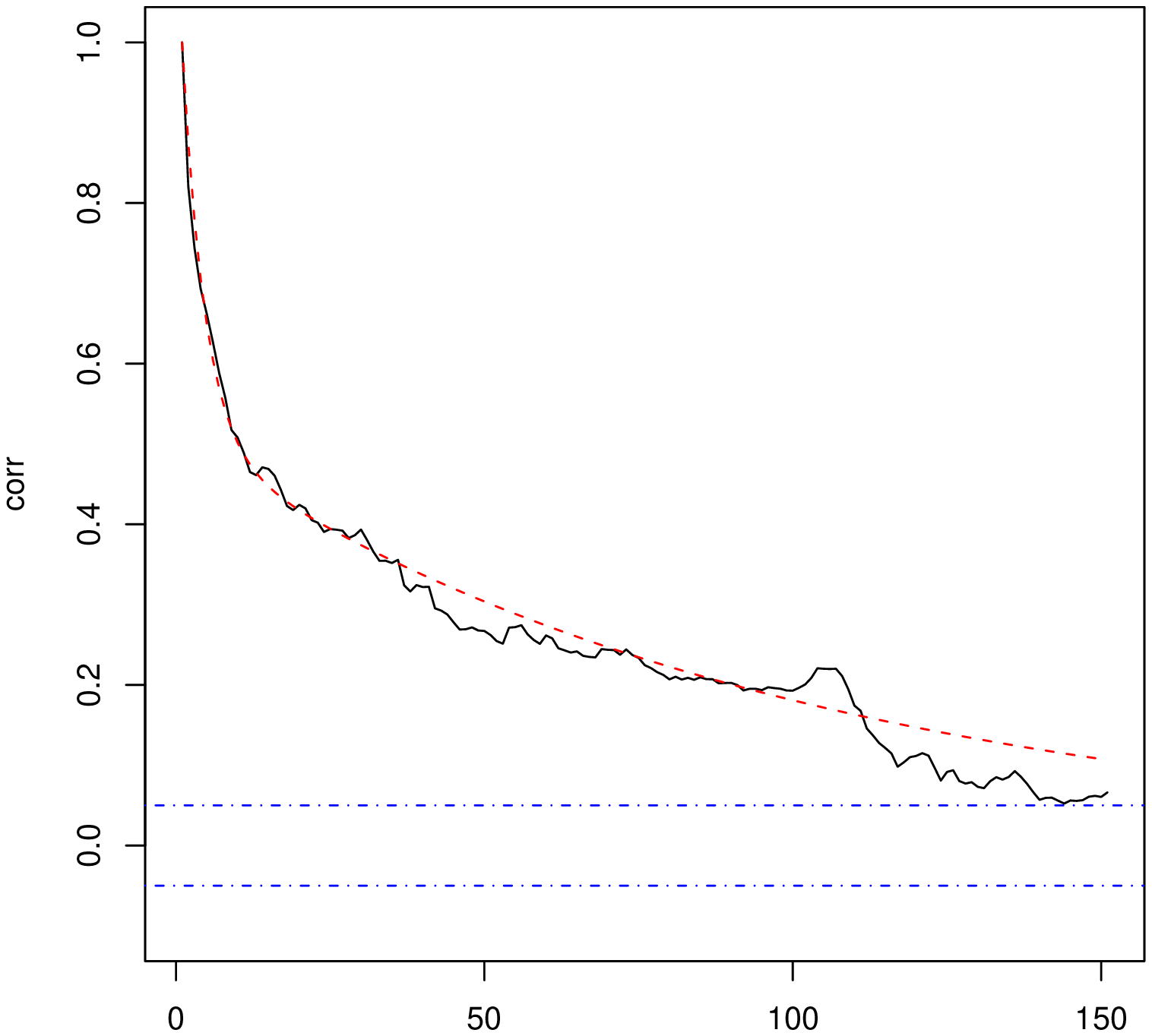}}
	\subfigure[CCF of $(Y^g,Y^e)$]{\includegraphics[width=5cm,height=5cm]{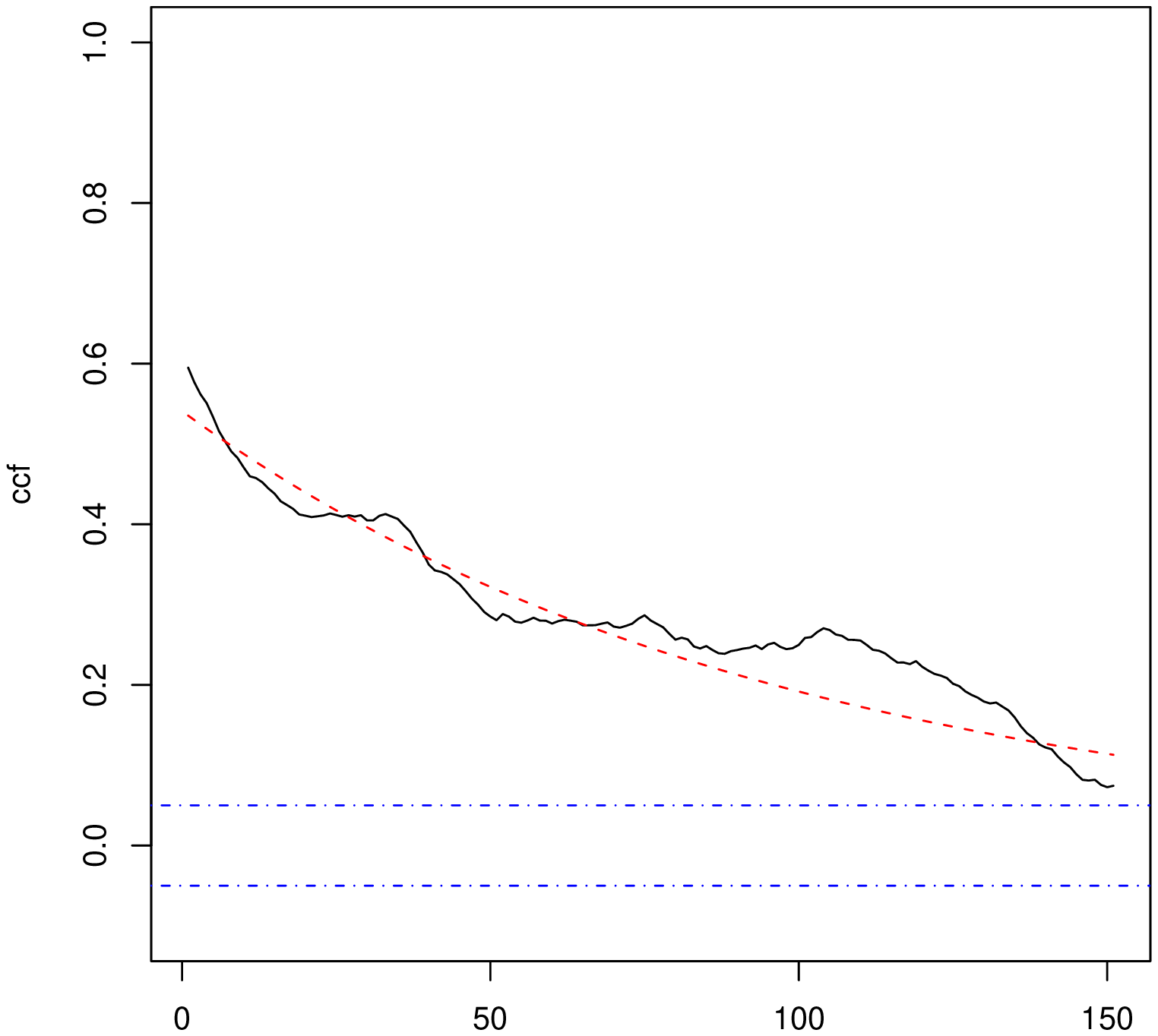}}
\caption{\label{fig-acfs} Empirical ACF and CCF of deseasonalized gas spot price and electricity spot price}
\end{center}
\end{figure}
Using a least squares approach, we fitted simulteanously $\rho^{g}(\tau)$, $\rho^{e}(\tau)$, $\rho^{g,e}(\tau)$ ($\tau=1,...,150$) to the empirical ACFs and CCF. We assumed that the observed spot prices have reached the stationarity. Both empirical and fitted ACFs\footnote{$\phi^{g}_{1}=0.43$, $\lambda^{g}_{1}=7.2$, and $\phi^{e}_{1}=0.49$, $\lambda^{e}_{1}=69.4$} and CCF\footnote{$\phi^{g,e}=0.53$, $\lambda^{g}_{2}=\lambda^{e}_{2}=\lambda^{g,e}=2.6$} are plotted in Figure \ref{fig-acfs}. 

We can see the separation into a fast speed of mean reversion for gas and electricity spot prices $\lambda^{g}_{1}$ and $\lambda^{e}_{1}$ which corresponds to a correlation dependence of approximately 2 and 30 days probably due to the spikes components whereas the slower speed of mean reversion corresponds to a correlation dependence of 64 days and corresponds to the stochastic equilibrium or the normal variation of gas and electricity spot prices.

\section{Theoretical background}
In order to modelize heavy tails (and spikes) of stationary spot prices distribution, a natural idea is to consider an ergodic diffusion process like representation of deseasonalized spot prices. 

In this section, we briefly recall how to construct a one dimensional process $X$ solution of a stochastic differential equation with a prespecified continuous invariant density $f$. Throughout the sequel we assume that $f$ is a strictly positive bounded continuous probability density on $(l,r)$ (and zero outside $(l,r)$). 

\subsection{The general case}
Let $\proc{X}$ the diffusion solution of the following stochastic differential equation (SDE)
\begin{equation*} \label{sde-dim1} \tag{$E_{b,\sigma}$}
	\d X_t = b(X_t) \d t + \sigma(X_t) \d B_t, \quad X_0 \in (l,r), 
\end{equation*}
where $b:(l,r) \rightarrow \R$ and $\sigma: (l,r) \rightarrow \R$ are locally Lipschitz functions and that $\sigma$ is not degenerate on $(l,r)$ \ie $\forall x \in (l,r), \sigma^2(x) > 0$. We introduce for the diffusion $\proc{X}$, the scale function $p:(l,r) \rightarrow \R$ defined for any $c \in (l,r)$ by 
\begin{equation*}
	\forall x \in (l, r), \quad p(x) = \int_c^x \exp \pa{- \int_c^y \frac{2 b(z)}{\sigma^2(z)} \d z} \d y, 
\end{equation*}
and the speed measure density $m:(l,r) \rightarrow \R_+^*$ defined by 
\begin{equation} \label{speed_m}
	\forall x \in (l, r), \quad m(x) = \frac{2}{p'(x) \sigma^2(x)} = \frac{2}{\sigma^2(x)} \exp\pa{\int_c^x \frac{2 b(z)}{\sigma^2(z)} \d z}.
\end{equation}
We recall (see e.g. \cite{karatzas-shreve,karlin-taylor}) that the process $\pa{p(X_t^\zeta)}_{t \ge 0}$ with $\zeta=\inf \ac{t \ge 0, X_t = l \text{ or } X_t = r}$ is a local martingale if and only if $p$ is the scale function (unique up to an affine transformation). Moreover, if the diffusion $\proc{X}$ is positive recurrent, the stationary probability distribution $\nu$ defined on $(l,r)$ satisfies 
\begin{equation*}
	\nu(\d x) = C m(x) \d x \quad \text{with} \quad  C = \pa{\int_l^r m(x) \d x}^{-1}.
\end{equation*}
This classical result is the key to construct a one-dimensional ergodic process that fits prescribed
stationary probability distribution. For a more general result to construct an inhomogeneous Markov martingale process that has prespecified marginal density we refer to \cite{madan-yor}.

\begin{prop}
Let $b:(l,r) \rightarrow \R$ be a continuous drift function. Suppose that $b$ and $f$ satisfy the following conditions 
\begin{equation} \label{hyp-B} \tag{$\mathcal{H}_B$}
	\quad \forall x \in (l,r), \; \int_l^x b(y) f(y) \d y > 0, \quad \text{and} \quad \int_l^r b(y) f(y) \d y = 0, \end{equation}
Then there exists a unique continuous diffusion function defined by 
\begin{equation*}
	\forall x \in (l,r), \quad \sigma(x) = \sqrt{2 \frac{\int_l^x b(y) f(y) \d y}{f(x)}},
\end{equation*}
such that \eqref{sde-dim1} has a unique solution $\proc{X}$, which is an ergodic diffusion process with stationary distribution $\nu$ satisfying $\nu(\d x) = f(x) \d x$.
\end{prop}

Further details of the proof outlined below can be found in \cite{sorensen}. 
\begin{proof}
Let $B$ be the function defined by $B(x) = \int_l^x b(y) f(y) \d y$. One checks easily that the scale function of $\proc{X}$ satisfies 
\begin{equation*}
	\forall x \in (l, r), \quad p(x) = B(c) \int_c^x \frac{1}{B(y)} \d y. 
\end{equation*}
One then obtains that $\lim_{x \rightarrow l} p(x) = -\infty$ and $\lim_{x \rightarrow r} p(x) = +\infty$. 

On the other hand, the speed measure of $\proc{X}$ has density $m$ that satisfies 
\begin{equation*}
	\forall x \in (l, r), \quad m(x) = \frac{f(x)}{B(x) p'(x)} = \frac{f(x)}{B(c)}.
\end{equation*}
The normalized speed measure density is then equal to the probability density $f$.

To prove existence and uniqueness of the solution $\proc{X}$, one proves existence and uniqueness of the process $\pa{p(X_t)}_{t \ge 0}$ satisfying a SDE without drift (see \cite{karatzas-shreve}).
\end{proof}

\begin{cor} Let $b: x \in (l,r) \mapsto - \lambda(x - \mu)$ and assume that probability density $f$ has expectation~$\mu$ and finite variance. Then there exists a unique continuous diffusion function defined by 
\begin{equation*}
	\forall x \in (l,r), \quad \sigma(x) = \sqrt{\frac{\int_l^x 2 \lambda (\mu - y) f(y) \d y}{f(x)}},
\end{equation*}
such that \eqref{sde-dim1} has a unique solution $\proc{X}$, which is an ergodic diffusion process with stationary distribution $\nu$ satisfying $\nu(\d x) = f(x) \d x$, and ACF given by 
\begin{equation*}
	\forall t, \tau \ge 0, \quad \corr(X_{t+\tau}, X_t) = e^{-\lambda \tau}. 
\end{equation*}
\end{cor}
The squared diffusion coefficients are explicitly known for a large number of commonly used probability diffusions. However, for some specific distributions, it is not possible to obtain a closed form of the diffusion coefficient. An approximation based on saddlepoint technique and the moment generating function (which is generaly known explicitly) is developed in \cite{sorensen}. 

\subsection{Quasi-Saddlepoint approximation}
We first recall that saddlepoint approximations are constructed by performing various operations on the moment generating function (MGF) of a random variable (see e.g. \cite{butler}). Let $X$ be an absolutely continuous random variable with density $f$ (with respect to the Lebesgue measure on $(l,r)$), moment generating function $M(t)$ and cumulant-generating function $\kappa(t) = \log M(t)$. Then the first-order saddlepoint density approximation to $f$ is given by
\begin{equation*}
	\forall x \in (l,r), \quad \hat f(x) = \pa{2 \pi \kappa''(\hat t_x)}^{-1/2} e^{-\pa{\hat t_x x - \kappa(\hat t_x)}},
\end{equation*}
where $t = \hat t_x$ is the (unique) solution to the saddlepoint equation $\kappa'(t) = x$, and primes denote derivatives.  We assume that the probability density $f$ has expectation $\mu$, \ie $\mu = \kappa'(0)$.

Considering the continuous differentiable function $\hat t: x \mapsto \hat t_x$, an integration by parts gives 
\begin{align*}
	\int_0^x \hat{t}(y) \d y & = \hat{t}(x) x - \int_0^x \hat{t}'(y) y \d y, \\
	& = \hat{t}(x) x - \int_0^x \d \kappa(\hat{t}(y)),
\end{align*}

\noindent since $y = \kappa'(\hat{t}(y))$. The saddlepoint density $\hat{f}$ writes then 
\begin{equation} \label{f_hat}
	\forall x \in (l,r), \quad \hat f(x) = \pa{2 \pi \kappa''(\hat t(x))}^{-1/2}  \exp\pa{-\int_0^x t(y) \d y}. 
\end{equation}
To construct an ergodic process $\proc{X}$ solution of \eqref{sde-dim1} with prespicified stationary density $\hat f$, the exponential terms that appear in \eqref{f_hat} and \eqref{speed_m} suggest the relation $\frac{-2 b}{\sigma^2} = t$. This construction is not exact but in \cite{sorensen} is proved that the speed density $m$ of $X$ is approximately propotional to the saddlepoint density $\hat f$. To be precise both $\sqrt{\kappa''(\hat t(x))}$ and $\sigma^2(x)$ are approximately proportional to $\kappa''(0) + \frac{1}{2} \kappa^{(3)}(0) \hat t(x)$ near the mean of the distribution. From now this normalized speed density $m$ will be called the \emph{quasi-saddlepoint} density approximation to $f$.

To summarize, if the saddlepoint function $\hat t$ is explicity known and efficiently computed, then we consider the diffusion with drift $b$, such that $b > 0$ on $(l,\mu)$ and $b < 0$ on $(\mu, r)$, and with diffusion coefficient 
\begin{equation*}
	\forall x \in (l,r), \quad \sigma(x) = \sqrt{\frac{-2 b(x)}{\hat t(x)}},
\end{equation*}
which is ergodic with stationary distribution $\tilde f(x) = \frac{c}{\sigma^2(x)} e^{-\pa{x \hat t(x) - \kappa(\hat t(x))}}$ (where $c$ is a normalizing factor), the quasi-saddlepoint density approximation to $f$ (see \cite{sorensen} Theorem 3.1 for more details).
 
The following example will become useful later when we are going to modelize deseasonalized gas and electricity spot prices.
\begin{example}\emph{The NIG-distribution}
The normal-inverse Gaussian (NIG) distribution is a member of the class of generalized hyperbolic distributions (see e.g. \cite{barndorff}). The NIG density is given by 
\begin{equation*}
	f(x)=\frac{\alpha\delta K_{1}\left(\alpha\sqrt{\delta^2+(x-l)^2}\right)}{\pi \sqrt{\delta^2+(x-l)^2}}\times e^{\delta \sqrt{\alpha^{2}-\beta^{2}} + \beta(x-l)}, \ \ x\in \mathbb{R},
\end{equation*}
where $\beta \in \R$, $\alpha > |\beta|$, $\delta > 0$, $l \in \mathbb{R}$ and $K_{1}$ is the the modified Bessel function of third order and index $1$. Note that if $X \sim \NIG\left( \alpha,\beta,\delta,l \right)$ then its two first moments are 
\begin{equation*}
	\esp{X} = l + \frac{\delta \beta}{\sqrt{\alpha^2 - \beta^2}} \quad \text{et} \quad \var(X) = \frac{\delta \alpha^2}{(\alpha^2 - \beta^2)^{\frac{3}{2}}}.
\end{equation*}
The two parameters $\delta$ and $l$ determine respectivelly the scale and the location of the law, and the two parameters $\alpha$ and $\beta$ determine the shape: $\alpha$ being responsible for the tail heavyness and $\beta$ for the skewness (asymmetry).

The cumulant-generating function $\kappa$ of the NIG distribution is defined for all $t$ such that $|\beta+t|<\alpha$ by 
\begin{equation*}
\kappa(t)= l t + \delta \left(\sqrt{\alpha^2-\beta^2}-\sqrt{\alpha^2-(\beta+t)^2}\right), 
\label{logMGF}
\end{equation*}
and the saddlepoint function is defined by 
\begin{equation*}
	\forall x \in \R, \quad \hat t(x) = \frac{\alpha\left(x-l\right)}{\sqrt{\delta^2+(x-l)^2}}-\beta.
\end{equation*}
In order to have an Ornstein process solution of \eqref{sde-dim1} with stationary density the quasi-saddlepoint density approximation $\tilde f$ to $f$, we consider the following drift and diffusion functions 
\begin{equation} \label{coef-b-s}
	\forall x \in \R, \quad b(x) = -\lambda \pa{x - \mu} \quad \text{and} \quad \sigma^2(x) = \frac{2 \lambda \sqrt{\delta^2+(x-l)^2} \pa{x-\mu}}{\alpha\left(x-l\right) - \beta \sqrt{\delta^2+(x-l)^2}},
\end{equation}
with $\mu = l + \frac{\delta \beta}{\sqrt{\alpha^2 - \beta^2}}$. 
\end{example}

\section{Cross-commodity multi-factor model}
In this section, we present two class of cross-commodity multi factor models: the geometric and the arithmetic class. Those two class are commonly used in stochastic modelling of commodity prices. The first one ensures the positivity of simulated spot prices. However, when dealing with forward contracts which have a delivery period or options pricing, the second one is analytically more tractable. Both class of models are based on stationary diffusion-type models analyzed in \cite{sorensen}. \subsection{Proposed modelization}

In this section, we propose an alternative model which captures the stylized features described in Section 2 without introducing jump diffusions. Sums of this kind of diffusions can fit the multi-scale ACFs and the CCF obtained for the deseasonalized gas and electricity spot prices.  

In order to represent the ACFs and CCF of gas and electricity deseasonalized spot prices, we are led to introduce stochastic processes that are sums of diffusions defined by $(E_{b,\sigma})$. To be more precise, we focus on the following two factor modelization for the deseasonalized log spot prices $Y^{g}$ and $Y^{e}$ 
\begin{equation}
Y^{g}_t = X^{g}_t + Z_t, \ \ \mbox{ and } \ \ Y^{e}_t = X^{e}_t + Z_t,
\label{underlyingprocesses}
\end{equation}
where $\proc{Z}$, $\proc{X^g}$ and $\proc{X^e}$ are mutually independant processes defined as following: 
\begin{itemize}
\item the process $\proc{Z}$ accounts for the stochastic equilibrium between both commodities with a slow rate of mean reversion $\lambda_{z}=\lambda_{2}^{g}=\lambda_{2}^{e}$. Thus, it represents the normal variation and will be defined by an Ornstein-Uhlenbeck process
\begin{equation}
\d Z_t = -\lambda_z Z_t \d t + \sigma_{z} \d W^z_t,
\label{diffugaussienne}
\end{equation}
with $\lambda_z > 0$ and $\sigma_z \in \R$. Note that $Z$ is ergodic with the Gaussian invariant probability $\mathcal{N}\left(0,\sigma_{z}^{2}/2\lambda_{z}\right)$.

\item the processes $\proc{X^{g}}$ and $\proc{X^{e}}$ represent the spikes component for each commodity. We modelize them by general Ornstein processes with high rate of mean reversion $\lambda_g = \lambda_1^g > 0$ and $\lambda_e = \lambda_2^e > 0$, namely
\begin{equation}
\d X^{j}_t = - \lambda_j \pa{X^{j}_t - \mu_{j}} \d t + \sigma_j(X^j_t; \theta_j) \d W^j_t, \ j=g,e,
\label{diffunig}
\end{equation}
where $\sigma_j$ is a parametric diffusion function such that $\proc{X^g}$ is an ergodic diffusion with invariant probability $f^j(.,\theta_j)$.
\end{itemize}

\begin{rem}
The following contruction can be extended to a more general multi-factor model. We can consider $m$ general Ornstein processes and $p$ Ornstein-Uhlenbeck processes so that
\begin{align*}
Y^{g}(t) & = \sum_{i=1}^{m} X^{g}_{i}(t) + \sum_{j=1}^{p} Z_{j}(t), \\
Y^{e}(t) & = \sum_{i=1}^{m} X^{e}_{i}(t) + \sum_{j=1}^{p} Z_{j}(t),
\end{align*}
where all processes are assumed to be mutually independent, i.e. driven by independent Wiener processes. We already observed that a two-factor model ($m=1$ and $p=1$) fits the ACFs and CCF well.
\end{rem}

\begin{prop}[The correlation structures]
Let $Y^{g},\ Y^{e}$ be the processes defined in \eqref{underlyingprocesses}. Then, the ACFs of $Y^{g}$ and $Y^{e}$ with lag $\tau>0$ are given by 
\begin{align*}
\rho^{g}(\tau)  =  \corr\pa{Y^{g}_{t+\tau},Y^{g}_t} &= \phi_{g} e^{-\lambda_g\tau} + (1-\phi_{g}) e^{-\lambda_{z}\tau},\\
\rho^{e}(\tau)  =  \corr\pa{Y^{e}_{t+\tau},Y^{e}_t} &= \phi_{e} e^{-\lambda_e\tau} + (1-\phi_{e}) e^{-\lambda_{z}\tau},
\label{ACF}
\end{align*}
where 
\begin{equation*}
\phi_{g}=\frac{\textnormal{Var}\left(X^{g}(t)\right)}{\textnormal{Var}\left(Y^{g}(t)\right)}, \  \ \mbox{ and } \ \  \phi_{e}=\frac{\textnormal{Var}\left(X^{e}(t)\right)}{\textnormal{Var}\left(Y^{e}(t)\right)}.
\end{equation*}
The CCF with lag $\tau>0$ is given by
\begin{equation*}
\rho^{g,e}(\tau):=\corr \pa{Y^{g}_{t+\tau},Y^{e}_t} = \phi_{g,e} \ e^{-\lambda_{z}\tau}, 
\label{ccfYgYe}
\end{equation*}
with,  $\phi_{g,e}=\frac{\textnormal{Var}\left(Z(t)\right)}{\sqrt{\textnormal{Var}\left(Y^{g}(t)\right)\textnormal{Var}\left(Y^{e}(t)\right)}}.$
\end{prop}

From the definition of $\phi_{g,e}$, we find that $\sigma_{z}^{2}=2\lambda_{z} \phi_{g,e} \sqrt{\textnormal{Var}\left(Y^{g}(t)\right)\textnormal{Var}\left(Y^{e}(t)\right)}$, where the last term is the product of the two stationary variance of the two processes. Consequently, one can easily derive $\sigma_{z}$ from the ACFs and CCF calibration.

\subsection{Calibration}\label{sec-calibration}
We propose a three-step calibration procedure for the model described above.

\subsubsection*{Step 1: Deseasonalizing spot prices}
We fit the seasonality functions $g(t)$ and $e(t)$ defined in section 2.1 to the logarithmic spot prices. The parameters of the functions are estimated using the least squares approach. Now, we focus on the deseasonalized spot prices $Y^{g}$ and $Y^{e}$ defined by
$$
Y^{g}(t)=\log\left(S^{g}(t)\right)-\log\left(g(t)\right) \ \ \mbox{ and } \ \ Y^{e}(t)=\log\left(S^{e}(t)\right)-\log\left(e(t)\right).
$$
One can consider the deseasonalized spot prices $e^{Y^{g}(t)}$ and $e^{Y^{e}(t)}$ instead of this geometric approach. 

\subsubsection*{Step 2: ACFs and CCF}
The least squares method consists in fitting the empirical ACFs $\rho^{g}(\tau)$, $\rho^{e}(\tau)$ and CCF $\rho^{g,e}(\tau)$ defined in section 2.4 to the empirical ones $\left(\tilde{\rho}^{g}(\tau)\right)_{\tau=1, ...,l}$, $\left(\tilde{\rho}^{e}(\tau)\right)_{\tau=1, ...,l}$, $\left(\tilde{\rho}^{g,e}(\tau)\right)_{\tau=1, ...,l}$ in order to derive the three speeds of mean reversion $\lambda_{1}^{g}$, $\lambda_{1}^{e}$, $\lambda_{z}$ with the diffusion coefficient $\sigma_{z}$ of the stochastic equilibrium process $Z$. This can be done by minimizing the sum of squared differences, namely 
$$
\argmin_{\lambda_g,\lambda_e,\lambda_z,\sigma_{z}} \sum_{\tau=1}^{l} \left(\left(\rho^{g}(\tau)-\tilde{\rho}^{g}(\tau)\right)^2+\left(\rho^{e}(\tau)-\tilde{\rho}^{e}(\tau)\right)^2+\left(\rho^{g,e}(\tau)-\tilde{\rho}^{g,e}(\tau)\right)^2\right).
$$
Stability tests showed that the estimates are robust with respect to small changes in the initial values of the parameters.
 
\subsubsection*{Step 3: Estimating the parameters of the spikes component}

The final step consists in statiscally estimating the parameters $\theta_g$ of the invariant density $f^{g}(.,\theta_g)$ of the process $X^{g}$ and the parameters $\theta_e$ of the invariant density $f^{e}(.,\theta_e)$ of the process $X^{e}$. For instance, if one decide to choose the quasi-saddlepoint approximation to the NIG density for $f^{g}$ and $f^{e}$, there will be four parameters to fit for each density. The model proposed is a sum of diffusion processes and hence is not Markovian. Thus, the likelihood cannot be written down explicitly. To overcome this problem, we use the maximum likelihood estimation of order $m$ ($m=0$ or $m=1$ in our case) method for stationary processes introduced in \cite{Azzalini1983}. Strong consistency and a Central Limit Theorem are proved for such estimates. It consists in approximating the log-likelihood of the serie $(y^{j}_{k})_{1 \leq k \leq n}$ ($j=g, \ e$), where $n$ is the number of sample points, by a sum whose generic term is the density function of $Y^{j}_{k}$ conditional on the $m$ most recent observations, for some $m\geq0$, namely,
\begin{equation}
\ell_{m}^{j}(\theta)=\sum_{k=1}^{n} \log\left(h^{j}(y^{j}_{k} \ | \ y^{j,m}_{k}; \theta_{j})\right),
\label{loglikelihood_order_m}
\end{equation}

\noindent where $y^{j,m}_{k}:=(y^{j}_{k-m}, \cdots, y^{j}_{k-1})$ and $h^{j}(.\ | \ y^{j,m}_{k}; \theta_{j})$ is the conditional probability density function of $Y^{j}_{k}$ given $Y^{j,m}_{k}=y^{j,m}_{k}$ for the parameters $\theta_{j}$. Note that if $m=0$, there is no conditioning and $h^{j}$ is simply the marginal density of $Y^{j}_{k}$, $k=1, \cdots, n$, which is the convolution of $Z_{k}$ and $X^{j}_{k}$. We suppose that $(X^{j}_{k})_{1\leq k \leq n}$ (resp. $(Z_{k})_{1\leq k \leq n}$) is ergodic with stationary distribution $f^{j}(.;\theta_{j})$ (resp. with Gaussian invariant probability $\mathcal{N}\left(0,\tilde{\sigma_{Z}}^{2}:=\sigma_{z}^{2}/2\lambda_{z}\right)$) so that the conditional probability density function is given by
\begin{equation}
h^{j}(y^{j}_{k};\theta_{j}) =\int_{-\infty}^{+\infty}
	f^{j}\left(y^{j}_{k}- \frac{\sigma_{Z}}{\sqrt{\lambda_{Z}}} u; \theta_j\right) \frac{e^{-u^2}}{\sqrt{\pi}} \d u, \quad j=g,e.
\label{density_order0}
\end{equation}

\noindent Note that it corresponds to the case of $(Y^{j}_{k})_{1\leq k \leq n}$ is independent and identically distributed random variables having the distribution of the stationary distribution of $Y^{j}$.

Numerically, the above integral can be approximated using a Gauss-Hermite quadrature method, namely 
\begin{equation*}
	h^{j}(y^{j}_{k};\theta_{j}) \approx \frac{1}{\sqrt{ \pi}} \sum_{k=1}^n 
	f^{j}\left(x- \frac{\sigma_{Z}}{\sqrt{\lambda_{Z}}} u_{k}; \theta_j\right) w_k, \quad j=g,e,
\label{produitconvapprox}
\end{equation*}

\noindent where ${(u_{k})}_{1 \le k\le n}$ are the roots of the Hermite polynomial $P_{n}$ and ${(w_{k})}_{1 \leq k \leq n}$ are the associated weights given by 
\begin{equation*}
	w_{k}=\frac{2^{n-1}n!\sqrt{\pi}}{n^{2}(P_{n-1}'(y_{k}))^2}, \quad k=1,...,n.
\end{equation*} 

If $m=1$, we need to compute the transition probability density $p_{Y^{j}_{k+1}|Y^{j}_{k}=y_{k}}(.; \theta_j)$ of $\proc{Y^{j}}$ for $j=g,e$. The two series $(X^{j}_{k})_{1 \leq k \leq n}$ and $(Z_{k})_{1 \leq k \leq n}$ are discrete observations of \eqref{diffunig} and \eqref{diffugaussienne}. Let $g$ be a Borel bounded test function and $k \in {1, \cdots, n-1}$, by conditioning we have
\begin{align*}
\mathbb{E}\left[g(Y^{j}_{k+1}) \ | \ Y^{j}_{k}=y_{k}\right] & = \int_{\mathbb{R}} \mathbb{E}\left[g(Y^{j}_{k+1}) \ | \ Y^{j}_{k}=y_{k}, \ Z_{k}=z\right] \mathbb{P}\left(Z_{k}=z \ | \ Y^{j}_{k}=y\right) \textnormal{d}z. \\
 & = \int_{\mathbb{R}} \mathbb{E}\left[g(X^{j}_{k+1}+Z_{k+1}) \ | \ Y^{j}_{k}=y_{k}, \ Z_{k}=z\right] \mathbb{P}\left(Z_{k}=z \ | \ Y^{j}_{k}=y\right) \textnormal{d}z.
\end{align*}

\noindent Note that the two processes $X^{j}$ and $Z$ are independent so that if we denote by $p_{X^{j}_{k}}(x^{j}_{k},.):=p_{X}(t_{k},t_{k+1}, x^{j}_{k}, .)$ and $p_{Z_{k}}(z_{k},.):=p_{Z}(t_{k},t_{k+1}, z_{k}, .)$, the conditional probability density functions of $X^{j}_{k+1}$ and $Z_{k+1}$ given $Z_{k}=z, \ X^{j}_{k}=x^{j}_{k}$, the expectation $\mathbb{E}\left[g(X^{j}_{k+1}+Z_{k+1}) \ | \ Y_{k}=y^{j}_{k}, \ Z_{k}=z\right]$ is given by 
$$
\int_{\mathbb{R}}  g(u) \int_{\mathbb{R}} p_{X^{j}_{k}}(y^{j}_{k}-z,v) p_{Z_{k}}(z,u-v)  \textnormal{d}v \textnormal{d}u.
$$


\noindent Moreover, we have
$$
\mathbb{P}\left(Z_{k}=z \ | \ Y^{j}_{k}=y^{j}_{k}\right)=\frac{\mathbb{P}\left(Z_{k}=z, \ Y^{j}_{k}=y^{j}_{k}\right)}{\mathbb{P}\left(Y^{j}_{k}=y^{j}_{k}\right)}= \frac{\mathbb{P}\left(Z_{k}=z\right) \mathbb{P}\left(X^{j}_{k}=y^{j}_{k}-z\right)}{\mathbb{P}\left(Y^{j}_{k}=y^{j}_{k}\right)},
$$

\noindent where,
$$
\mathbb{P}\left(Y^{j}_{k}=y^{j}_{k}\right)= \int_{-\infty}^{+\infty}
	f^{j}\left(y^{j}_{k}- u; \theta_j\right) \frac{1}{\sqrt{2\pi} \tilde{\sigma}_{Z}}e^{-\frac{1}{2 \tilde{\sigma}_{Z}^{2}} u^2} \d u, 
$$

\noindent and $\mathbb{P}\left(X^{j}_{k}=y^{j}_{k}-z\right)= f^{j}\left(y^{j}_{k}-z;\theta_{j}\right)$, $\mathbb{P}\left(Z_{k}=z\right)=\frac{1}{\sqrt{2\pi} \tilde{\sigma}_{Z}}e^{-\frac{1}{2\tilde{\sigma}_{Z}^{2}}z^{2}}$. Finally, one easily identifies the transition probability density $p_{Y^{j}_{k+1}|Y^{j}_{k}=y^{j}_{k}}(y; \theta_j)$ which is given by
$$
\int_{\mathbb{R}} \left(\int_{\mathbb{R}} p_{X^{j}_{k}}(y^{j}_{k}-z,v) p_{Z_{k}}(z,u-v)\textnormal{d}v\right) \frac{1}{\sqrt{2\pi}\tilde{\sigma}_{Z}}\frac{f^{j}\left(y^{j}_{k}-z;\theta_{j}\right) e^{-\frac{1}{2\tilde{\sigma}_{Z}^{2}}z^{2}}}{\mathbb{P}\left(Y^{j}_{k}=y^{j}_{k}\right)}\textnormal{d}u .
$$

\noindent Note that we have $p_{Z_{k}}(z_{k},z)=\frac{1}{\sqrt{2\pi} \bar{\sigma}_{Z}}e^{-\frac{1}{2 \bar{\sigma}_{Z}^{2}}z^{2}}$, with $\bar{\sigma}_{Z}=\sigma_{Z}\sqrt{\frac{1-e^{-2 \lambda_{Z}\Delta}}{2 \lambda_{Z}}}$ using an exact scheme of the Ornstein-Uhlenbeck process $(Z_{k})_{1 \leq k \leq n}$ of step $\Delta > 0$, namely
\begin{align}
Z_{k+1} & = e^{-\lambda_{z} \Delta} Z_{k} + 
\sigma_Z \sqrt{\frac{1-e^{-2 \lambda_Z \Delta}}{2 \lambda_Z}} G^z_{k+1}, \label{schemeulerz}
\end{align}

\noindent where $\left(G^z_{k}\right)_{k \geq 1}$ is a sequence of i.i.d. standard normal random variables. However, in most cases, there is no closed expression for $p_{X^{j}_{k}}(x^{j}_k,.)$. To overcome this problem one solution is to consider the transition probability density function $p_{\bar{X}^{j}_{k}}(x^{j}_k,.)$ of the Euler scheme $\left(\bar{X}^{j}_{k}\right)_{k \geq 0}$
\begin{align}
\bar{X}^j_{k+1} & = e^{-\lambda_{j} \Delta} \bar{X}^j_{k} + \mu_{j}\left(1-e^{-\lambda_{j} \Delta}\right)+
\sigma_j\left(\bar X^j_{k}; \theta_j\right) \sqrt{\frac{1-e^{-2 \lambda_j \Delta}}{2 \lambda_j}} G^j_{k+1}, \ k\geq0 \label{schemeulerx}
\end{align}

\noindent where $\left(G^j_{k}\right)_{k \geq 1}$ is a sequence of i.i.d. standard normal random variables independent of $\left(G^z_{k}\right)_{k \geq 1}$. Consequently, $p_{\bar{X}^{j}_{k}}(x^{j}_k,.)=\frac{1}{\sqrt{2\pi} \bar{\sigma}_{j}(x^{j}_{k};\theta_j)} e^{-\frac{1}{2 \bar{\sigma}^{2}_{j}(x^{j}_{k};\theta_j)}x^{2}}$, with $\bar{\sigma}_{j}(x^{j}_{k};\theta_j)=\sigma_j\left(\bar x^j_{k}; \theta_j\right)\sqrt{\frac{1-e^{-2 \lambda_{j}\Delta}}{2 \lambda_{j}}}$ so that we have
$$
\int_{\mathbb{R}} p_{\bar{X}^{j}_{k}}(y^{j}_{k}-z,v) p_{Z_{k}}(z,u-v)\textnormal{d}v=\frac{1}{\sqrt{2 \pi} \tilde{\sigma}(y^{j}_{k},z; \theta^{j})} e^{-\frac{1}{2\tilde{\sigma}^{2}(y^{j}_{k},z; \theta^{j})}\left(u-m^{j}_{k}\right)^{2}},
$$

\noindent where for $k\in \left\{1, \cdots, n \right\}$, $m^{j}_k=e^{-\lambda_Z \Delta}z+e^{-\lambda_{j} \Delta}(y^{j}_{k}-z)+\mu_{j}(1-e^{-\lambda_{j} \Delta})$ and $\tilde{\sigma}^{2}(y^{j}_{k},z; \theta^{j})= \bar{\sigma}_{j}^{2}\left(y^{j}_{k}-z; \theta_{j}\right) +\bar{\sigma}_{Z}^{2}$. 
\begin{rem}
In \cite{Elerian1998}, a transition probability density function based on Milstein scheme is used. In \cite{Kessler1997}, a gaussian transition probability density function with Taylor expansions is used to propose an efficient estimator for $\theta_j$.
\end{rem}

The method of maximum likelihood of order $m$ estimates $\hat{\theta}_{j,m}$ by finding the value of $\theta_{j}$ that maximizes \eqref{loglikelihood_order_m} using standard numerical optimization procedure.

\section{Simulation and application} \label{sec-sim-app}
\subsection{Empirical results on Powernext and NBP spot prices} \label{emp-res}
In this section, we perform the calibration procedure on electricity spot prices coming from the Powernext market and on gas spot prices at the NBP. Then, we perform a simulation with estimated parameters over the same period. To avoid negative prices, we choose to represent spot prices by an arithmetic model, namely
\begin{align}
S^{g}(t) &= g(t) \times e^{X^{g}(t)+Z(t)},  \label{Sg}\\
S^{e}(t) &= e(t) \times e^{X^{e}(t)+Z(t)},  \label{Se}
\end{align}
where $g(t)$, $e(t)$ are the trend and seasonality functions defined in Section \ref{sect-seaso}, $X^{g}$, $X^{e}$ are solutions of \eqref{sde-dim1} with $b$ and $\sigma$ defined in \eqref{coef-b-s} and $Z$ is a Gaussian Ornstein-Uhlenbeck	process solution of \eqref{diffugaussienne}. 

We choose the NIG distribution for those two processes in order to capture the heavy tails behavior observed on data, \emph{i.e.} large values with low probability that cannot be obtained by a Gaussian process. We observed that the quasi-saddlepoint approximation of the NIG-distribution is well suited to represent the two spike components. One can choose another distribution and devise the same calibration process as in the previous section. The results of steps 1 and 2 of the calibration procedure are reported in Figure \ref{fig-saiso} and the quality of the ACFs and CCF fits is represented in Figure \ref{fig-acfs}. Now, we proceed to the estimation of the four parameters $\theta_{g}=(\alpha_{g},\beta_{g},\delta_{g},l_{g})$ of the process $X^{g}$ and the four parameters $\theta_{e}=(\alpha_{e},\beta_{e}, \delta_{e}, l_{e})$ of the process $X^{e}$ using the maximum likelihood estimation method described in the previous section on the deseasonalized spot prices. We observed that the maximum likelihood estimation method of order 0\footnote{Fitted parameters of order 0 are: $\alpha_{g}= 1.93$, $\beta_{g}=0.90$, $\delta_{g}=2.25 e-3$, $l_{g}=-8.8e-3$ and $\alpha_{e}= 3.49$, $\beta_{e}=1.24$, $\delta_{e}= 0.08$, $l_{e}=0.11$.} is more robust and gives better results than the one of order 1\footnote{Fitted parameters of order 1 are: $\alpha_{g}= 0.76$, $\beta_{g}=7.8e-2$, $\delta_{g}=7.8e-4$, $l_{g}=-0.11$ and $\alpha_{e}= 1.56$, $\beta_{e}=0.34$, $\delta_{e}= 1.1 e-2$, $l_{e}=0.16$}. The initial parameters are set to $(1,0,1,0)$ for both components.

\begin{figure}[!ht]
\begin{center}
\includegraphics[width=7cm,height=6cm]{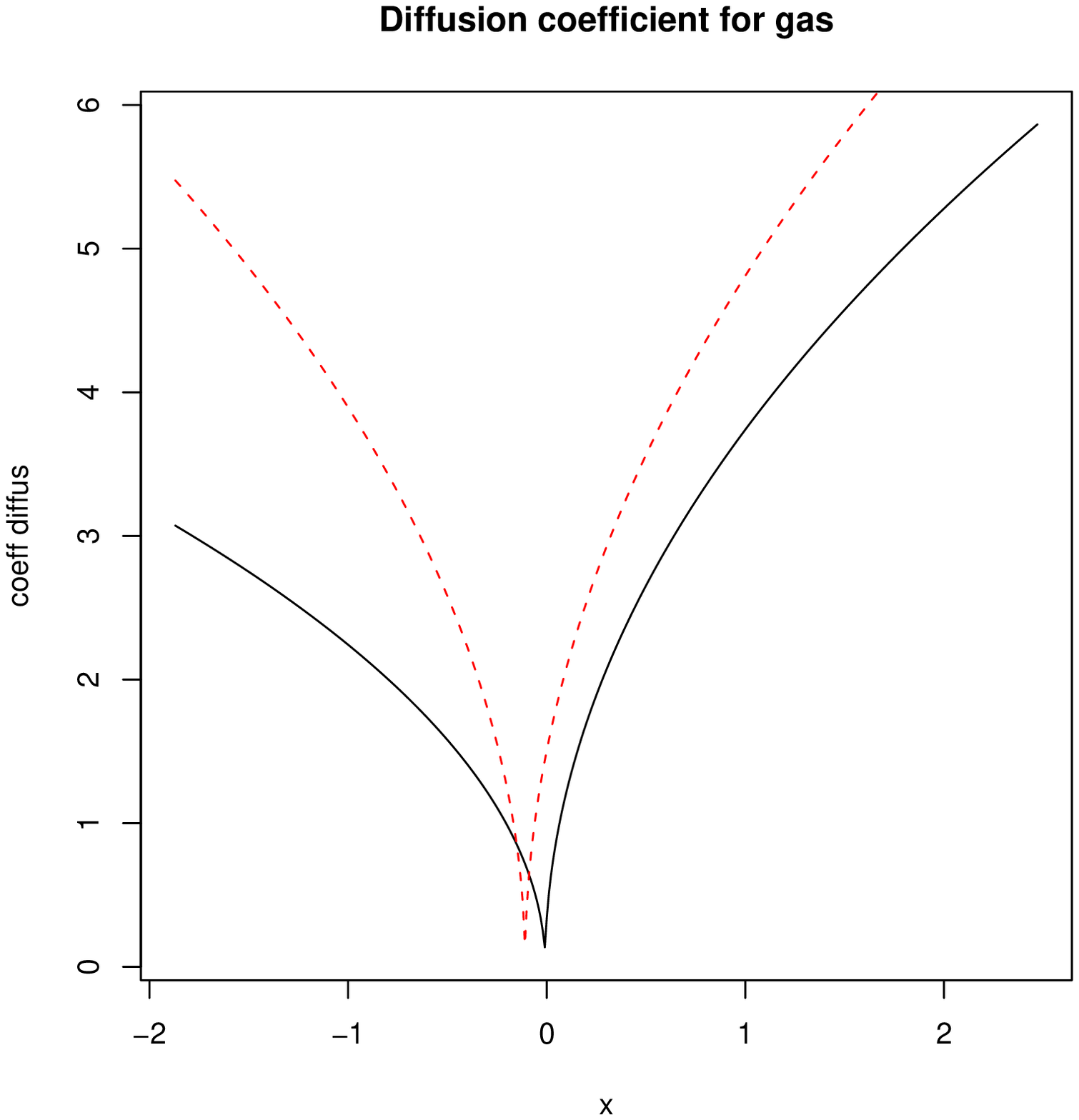}
\includegraphics[width=7cm,height=6cm]{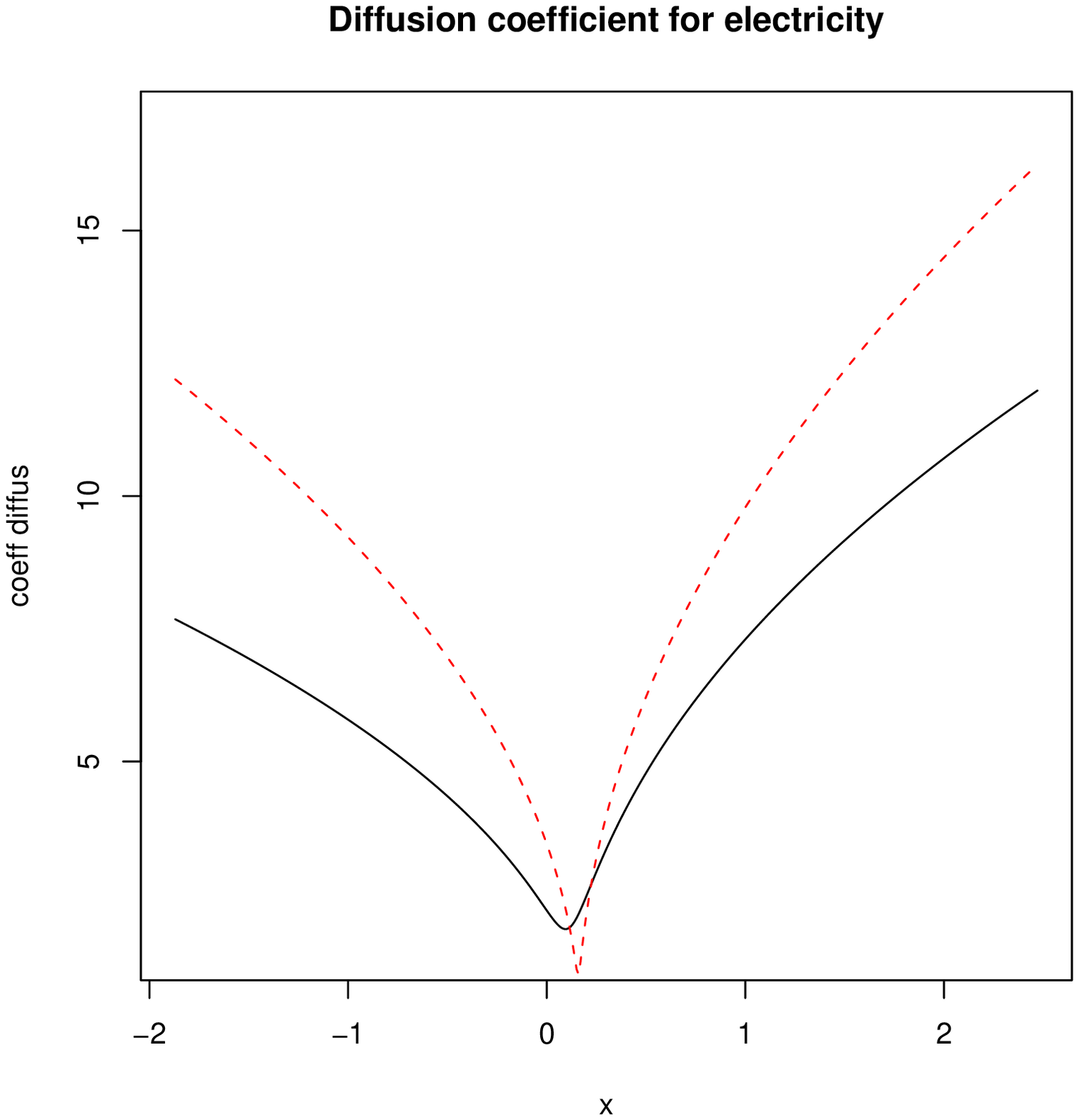}
\caption{\label{fig-sigma}
Squared diffusion coefficients using fitted parameters with maximum likelihood estimation of order 0 (normal lines) and of order 1 (dashed lines).}
\end{center}
\end{figure}

The algorithms converged quickly. The diffusion coefficient functions $\tilde{\sigma_{j}}(.,\theta_{j})$, $j=g, e$, with the fitted parameters, are documented in Figure \ref{fig-sigma}. We see that the shape of the diffuion coefficients are quite similar for the gas and electricity spot deseasonalized spot prices. Spikes are obtained when the processes $Y^{g}$ and $Y^{e}$ are far from their mean by clusters of volatility, \emph{i.e.} periods of high volatility. As we see, large values are more likely and the asymmetry is more pronounced for electricity spot prices than for gas spot prices. We clearly see spikes as cluster of volatility are more probable and more intense for electricity deseasonalized spot prices than for gas deseasonalized spot prices.

In order to simulate price trajectories, we consider the Euler-Maruyama schemes defined by \eqref{schemeulerx} and \eqref{schemeulerz}. If one is concerned by estimating some quantities (for instance quantiles) on only one trajectory then one should replace the above Euler schemes of $X^{g}$ and $X^{e}$ with their respective Milstein schemes $\tilde{X}^{g}$ and $\tilde{X}^{e}$ in order to achieve a smaller strong error rate. It consists in devising the following schemes for $j=g,e$,
\begin{multline*}
\tilde{X}^{j}_{t_{k+1}} = e^{-\lambda_{j}(t_{k+1}-t_{k})}\left(\tilde{X}^{j}_{t_{k}}+\left(\mu_{j}\lambda_{j}-\frac{1}{2}{\sigma_{j}}{\sigma_{j}}^{'}(\tilde{X}^{j}_{t_{k}}; \theta_{j})\right)\Delta\right)  \\
+\sigma_j\left(\bar X^j_{t_k}; \theta_j\right) \sqrt{\frac{1-e^{-2 \lambda_j \Delta}}{2 \lambda_j}} G^j_{k+1}+ \frac{1}{2}{\sigma_{j}}{\sigma_{j}}^{'}(\tilde{X}^{j}_{t_{k}}; \theta_{j}) \left(G^j_{k+1}\right)^{2}, \  \tilde{X}^{j}_{0}=x^{j}_{0},
\end{multline*}
where ${\sigma_{j}}^{'}$ is the first derivative of ${\sigma_{j}}$.    

\begin{figure}[!ht]
\begin{center}
\includegraphics[width=15cm,height=5.5cm]{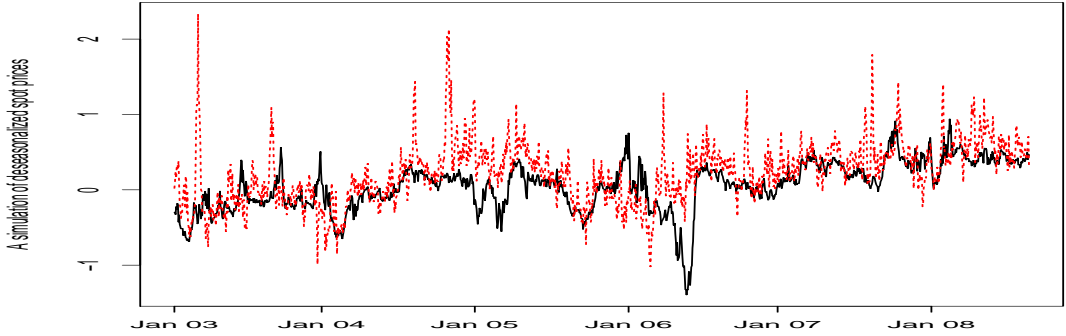}
\caption{\label{fig-sim-des} A simulation of gas (normal line) and electricity (dotted line) deseasonalized spot prices.}
\end{center}
\end{figure}

\begin{figure}[!ht]
\begin{center}
\includegraphics[width=7cm,height=6cm]{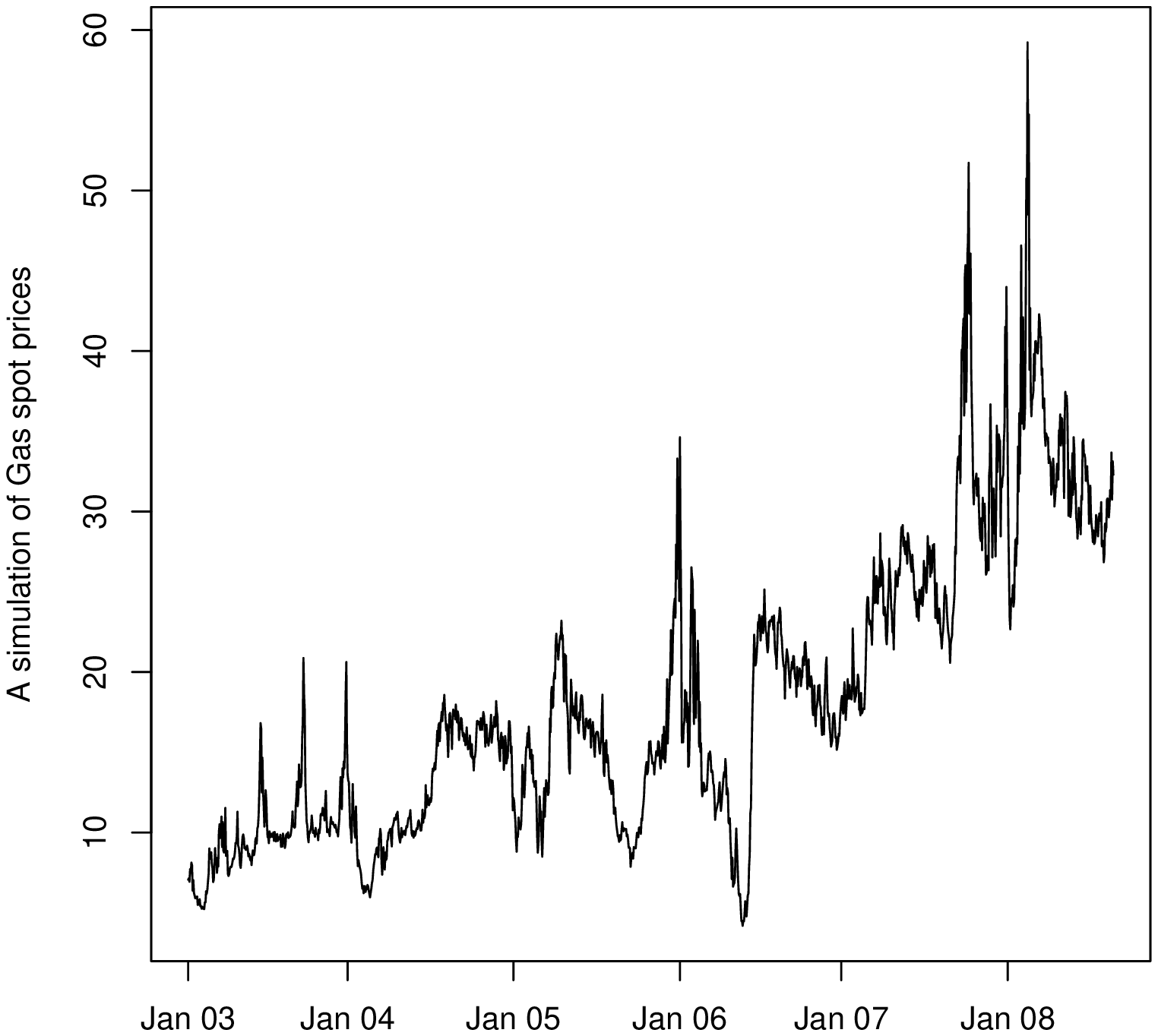}
\includegraphics[width=7cm,height=6cm]{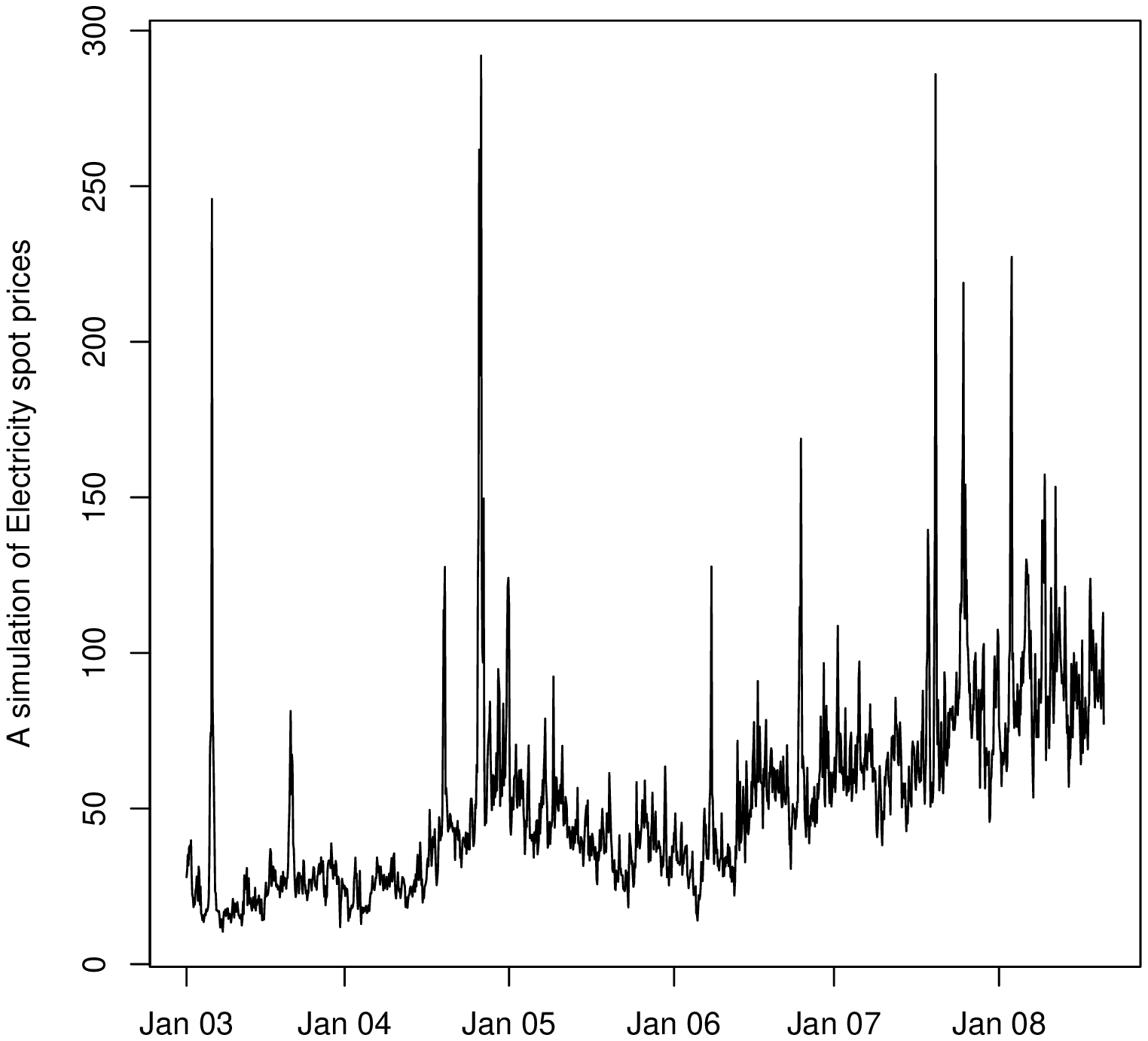}
\caption{\label{fig-sim-prix} Simulated Electricity spot prices on the Powernext market on the left and Gas spot prices at the NBP on the right for the period 14 January 2003 till 20 August 2008.}
\end{center}
\end{figure}

\begin{figure}[!ht]
\begin{center}
	\subfigure[ACF of one simulation of $Y^g$]{\includegraphics[width=7cm,height=6cm]{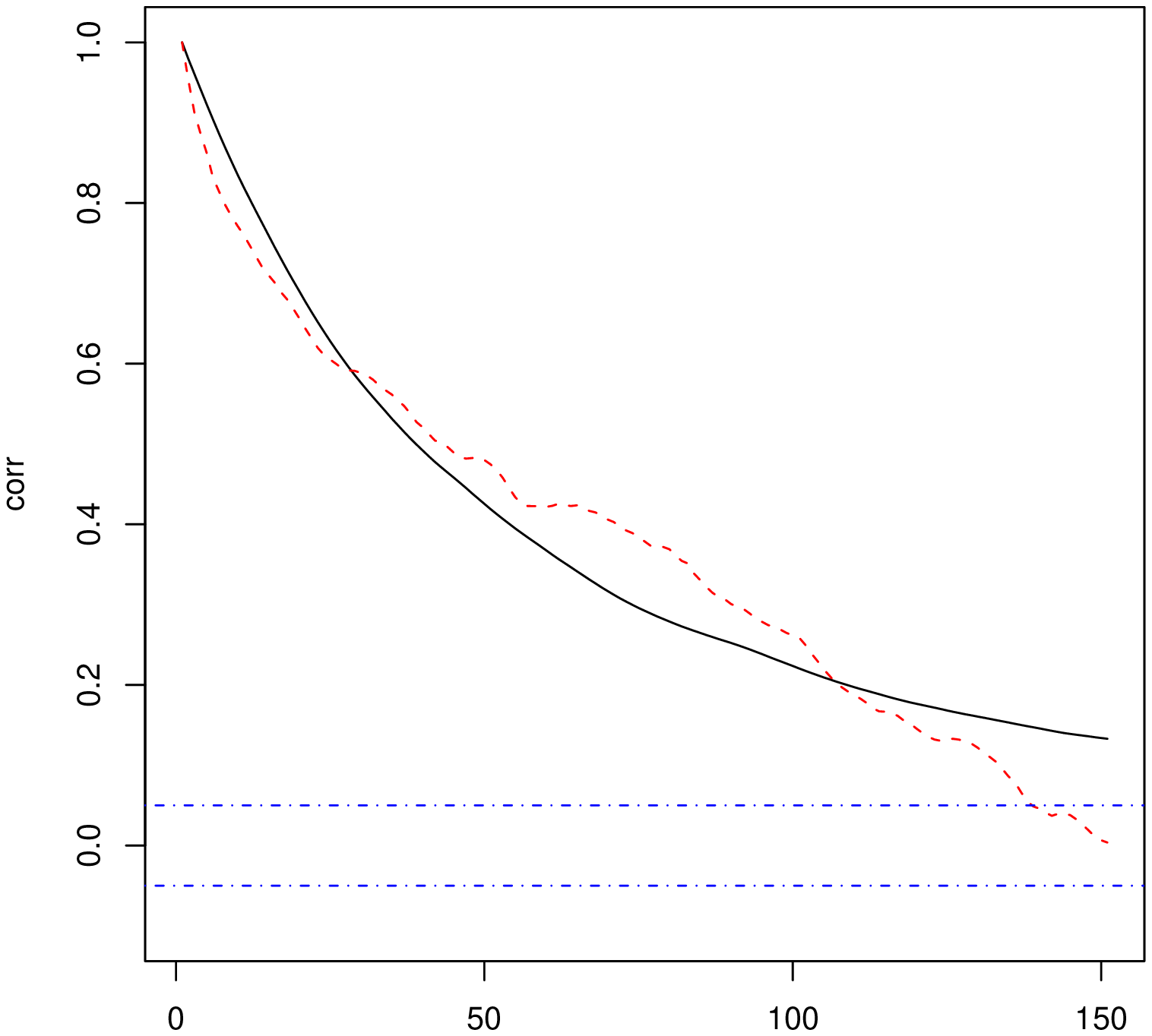}}
	\subfigure[ACF of one simulation of $Y^e$]{\includegraphics[width=7cm,height=6cm]{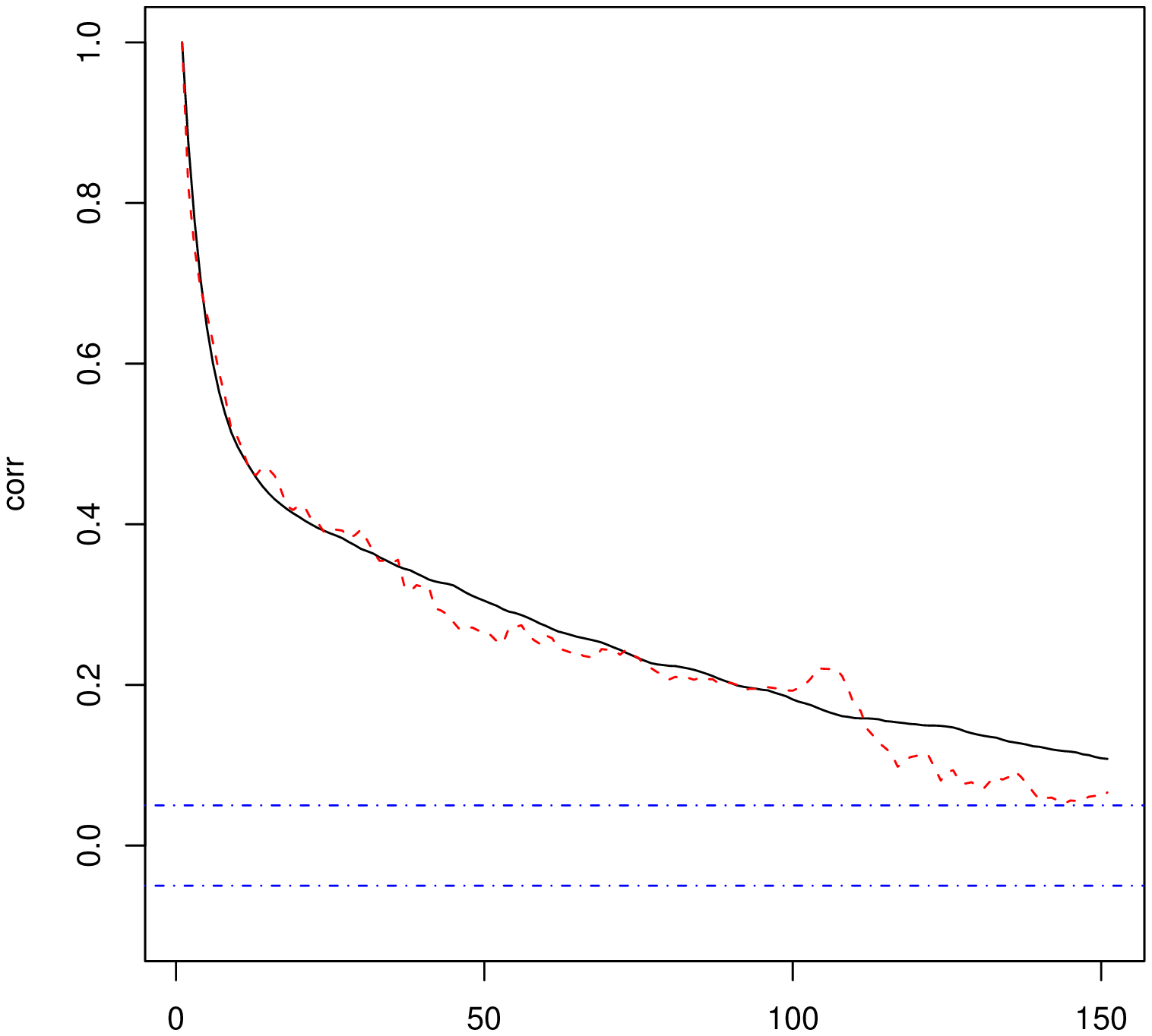}}
	\subfigure[CCF of the simulations]{\includegraphics[width=7cm,height=6cm]{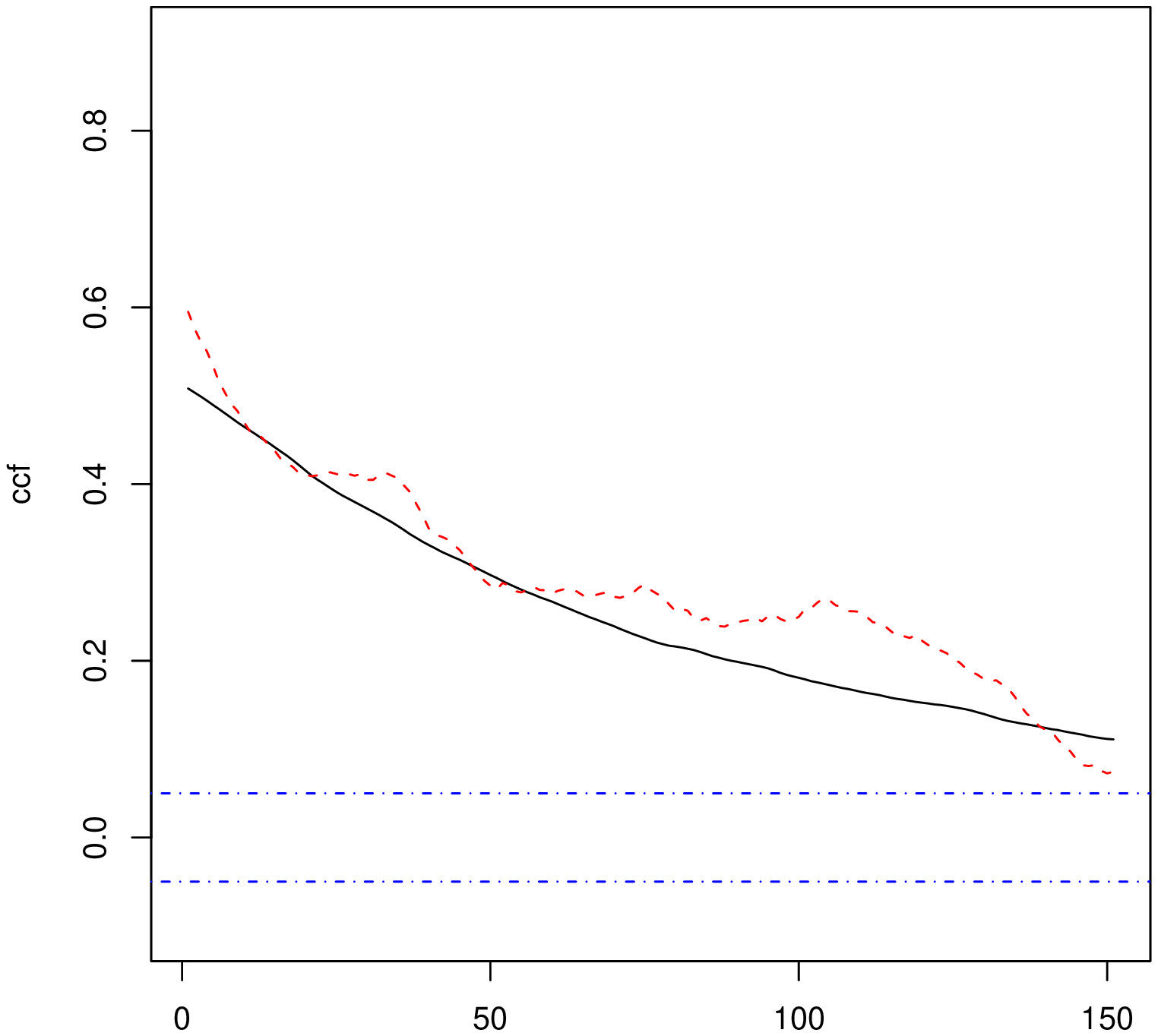}}
\caption{\label{fig-sim-acfs} ACFs and CCF of simulated gas and electricity spot prices (normal lines) with the historical ACFs and CCF (dotted lines).}
\end{center}
\end{figure}

In the following simulations, we consider Milstein schemes of step $t_{k}=k \Delta$, with $\Delta=\frac{1}{252}$. Next, we add to the simulated processes the two seasonality functions. In Figure \ref{fig-sim-des}, the simulated deseasonalized spot prices are represented. We see that both commodities are strongly linked and that the model mimics the statistical behaviour of the deseasonalized spot prices. In Figure \ref{fig-sim-prix}, the simulated spot prices are represented. In Figure \ref{fig-sim-acfs}, both simulated and historical ACFs and CCF are plotted. We clearly see that the model reproduces the correlation structures.

%

\subsection{Application: measuring risk of a cross-commodity portfolio}
In this section, we aim at measuring the risk of a portfolio composed of a short position in a power plant that produces electricity from gas day by day $t_{1} < \ ... < \ t_{N}$ for several maturities $T=t_{N}= \ 6$ months, $1$ year and $3$ years. The loss at time 0 of the portfolio with a time horizon $T$ can be written 
$$
L_{T}= \sum_{k=1}^{N} e^{-r t_{k}}\left(S^{e}_{t_{k}}-h_{R}S^{g}_{t_{k}}-C\right)_{+}-P_{T}^{c},
$$
where $r=5\%$ is the annual interest rate, $h_{R}=3$ denotes the Heat Rate, $C=5$ \euro/\textnormal{MWh} denotes the generation costs and where $P_{T}^{c}$ is an estimation of the price of the option on the power plant obtained by a crude Monte Carlo simulation, namely
$$
P_{T}^{c} \approx \sum_{k=1}^{N} e^{-r t_{k}}\esp{\left(S^{e}_{t_{k}}-h_{R}S^{g}_{t_{k}}-C\right)_{+}}.
$$

Since gas and electricity markets are incomplete, we price and estimate risk measures under the historical probability. In order to measure the risk, we consider the Value-at-Risk (VaR), which is certainly the most commonly used risk measures in the context of risk management. By definition, the Value-at-Risk at level $\alpha \in (0,1)$ (VaR$_{\alpha}$) of a given portfolio is the lowest amount not exceeded by its loss with probability $\alpha$. In this example, we set $\alpha=95\%$. Actually, for the considered portfolio, the VaR$_{\alpha}$ is the unique solution $\xi$ of the equation
$$
\prob{L_{T} \leq \xi} = \alpha.
$$
The portfolio's VaR$_{\alpha}$ is just a quantile of its loss and is interpreted as a reasonable worst case level. 

Now, we are interested in measuring the impact of the proposed model for gas and electricity spot prices on the portfolio's VaR. In order to do that, we consider three different models:
\begin{itemize}
	\item Case 1: the mean-reverting cross-commodity model (in its geometric form) proposed in this paper and defined by \eqref{Sg} and \eqref{Se}. It modelizes typical features of gas and electricity spot prices likes spikes and the long term dependency. 

\item Case 2: a slight modification of the previous model in which we do not take into account the dependence of the two energy spot prices. To be more precise, we consider the following model specification
\begin{align*}
S^{g}(t) & = g(t) \times e^{X^{g}(t)+Z^{g}(t)}, \\
S^{e}(t) & = e(t) \times e^{X^{e}(t)+Z^{e}(t)}, 
\end{align*}
where $X^{g}$ and $X^{e}$ are solutions of \eqref{sde-dim1} with $b$ and $\sigma$ defined in \eqref{coef-b-s}, and where $Z^{g}$, $Z^{e}$ are two independant Gaussian OU	processes solution of \eqref{diffugaussienne}. By this model, we want to measure the impact on the VaR$_{\alpha}$ of the long term dependency modeling. The calibration process is slightly modified since $S^{g}$ and $S^{e}$ are now independent. The step 2 is replaced by two different minimizations corresponding to each ACF. Steps 1 and 3 remain unchanged. 

\item Case 3: a slight modification of the case 1 in which we do not modelize the spikes feature. To be more precise, we replace the NIG-distributed processes by Gaussian Ornstein-Uhlenbeck processes, namely 
\begin{align*}
S^{g}(t) & = g(t) \times e^{Z^{g}(t)+Z(t)}, \\
S^{e}(t) & = e(t) \times e^{Z^{e}(t)+Z(t)}, 
\end{align*}
where $Z^{g}$, $Z^{e}$, $Z$ are three different Gaussian OU	processes solution of \eqref{diffugaussienne}. By this model, we want to quantify the impact on the VaR$_{\alpha}$ of the spike feature of gas and electricity spot prices.   
\end{itemize}

In each case, we estimate $P_{T}^{c}$ and the VaR$_{\alpha}$ using 10 000 Monte Carlo simulations. We devise Euler schemes of step $t_{k}=k \Delta$ with $\Delta=\frac{1}{252}$. In order to estimate the VaR$_{\alpha}$, we use the inversion of the simulated empirical distribution function. 

\begin{rem}
Since gas and electricity spot prices are sums of diffusion processes solution of $\left(E_{b,\sigma}\right)$, one can easily use the method investigated in \cite{frikha} to estimate the VaR$_{\alpha}$ and other risk measures. It is based on stochastic approximation algorithms with an adaptive variance reduction tool (unconstrained importance sampling algorithm). The method is known to achieve good variance reduction when $\alpha \approx 1$ as it is often the case. For the sake of simplicity, we only considered the classical method based on the inversion of the empirical distribution function.
\end{rem}

The results are summarized in Tables \ref{tab-res}. Note that for each case, the estimations are computed using the same pseudo-random number generator initialized with the same \emph{seed}. The number in parentheses refers to the $95\%$ confidence level. 
\begin{table}
\centering
\begin{tabular}{l | p{2cm} | r r | r}
& Maturity & $P_{T}^{c}\;\;$ &$(\pm$Error) & {VaR$_{\alpha}\;$}  \\
\hline 
\multirow{3}{3.7cm}{Case 1 \\ (Proposed model)}
& 6 months & 83.3   & ($\pm$3.3)   & 262.4 \\
& 1 year   & 220.1  & ($\pm$5.5)   & 495.4 \\
& 3 years  & 745.0  & ($\pm$11.2)     & 1081.0 \\
\hline
\multirow{3}{3.7cm}{Case 2 \\ (No cross-correlation)}
& 6 months & 51.2   & ($\pm$2.9)   & 250.1  \\
& 1 year   & 222.6  & ($\pm$8.4)   & 880.2   \\
& 3 years  & 850.6  & ($\pm$21.3)  & 2213.1  \\
\hline 
\multirow{3}{3.7cm}{Case 3 \\ (Gaussian model)}
& 6 months & 32.9  & ($\pm$1.1)  & 107.7 \\
& 1 year   & 129.8 & ($\pm$2.7)  & 275.9 \\
& 3 years  & 437.1 & ($\pm$5.8)  & 565.5 \\
\end{tabular}
\caption{Estimation of the price of the Power plant and the VaR$_{\alpha}$ of the portfolio.}
\label{tab-res}
\end{table}

We observe that there are slight differences in terms of the price $P_{T}^{c}$ between the case 1 and 2 but huge differences in terms of risk. Taking into account the long term correlation between gas and electricity spot prices can reduce substantially the risk of this portfolio. Modeling independently each energy spot prices leads to an overestimation of the VaR$_{\alpha}$ of the portfolio's loss. The results obtained by using the model investigated in case 3 shows that introducing the spikes behavior into the model can increase greatly both $P_{T}^{c}$ and the risk of the portfolio. We also estimated the same quantities using the arithmetic version of the three models presented above. We obviously obtained different values from the ones presented but the same conclusions hold: modeling adequatly the cross correlation between gas and electricity spot prices reduces the risk of portfolio whereas modeling adequatly the spiky behavior of both commodities increases greatly the price of the option and the risk associated to the portfolio.

\bibliography{biblio}
\bibliographystyle{abbrv}
\end{document}